\documentclass[journal]{IEEEtran}
\usepackage{amsmath,amsfonts}
\usepackage{algorithmic}
\usepackage{algorithm}
\usepackage{array}
\usepackage{textcomp}
\usepackage{stfloats}
\usepackage{url}
\usepackage{verbatim}
\usepackage{graphicx}
\usepackage{cite}
\usepackage{url}
\usepackage[hidelinks]{hyperref}  
\usepackage{xcolor}
\usepackage{acronym}
\usepackage{siunitx}
\usepackage{subcaption}
\usepackage{comment}
\usepackage{lipsum}
\usepackage{tfrupee} 
\usepackage{tabularx}
\newacro{IoT}[IoT]{internet-of-things}
\newacro{AQMS}[AQMS]{air quality monitoring stations}

\usepackage{scalerel}
\usepackage{tikz}
\usetikzlibrary{svg.path}

\definecolor{orcidlogocol}{HTML}{A6CE39}
\tikzset{
    orcidlogo/.pic={
        \fill[orcidlogocol] svg{M256,128c0,70.7-57.3,128-128,128C57.3,256,0,198.7,0,128C0,57.3,57.3,0,128,0C198.7,0,256,57.3,256,128z};
        \fill[white] svg{M86.3,186.2H70.9V79.1h15.4v48.4V186.2z}
        svg{M108.9,79.1h41.6c39.6,0,57,28.3,57,53.6c0,27.5-21.5,53.6-56.8,53.6h-41.8V79.1z M124.3,172.4h24.5c34.9,0,42.9-26.5,42.9-39.7c0-21.5-13.7-39.7-43.7-39.7h-23.7V172.4z}
        svg{M88.7,56.8c0,5.5-4.5,10.1-10.1,10.1c-5.6,0-10.1-4.6-10.1-10.1c0-5.6,4.5-10.1,10.1-10.1C84.2,46.7,88.7,51.3,88.7,56.8z};
    }
}

\newcommand\orcidicon[1]{\href{https://orcid.org/#1}{\mbox{\scalerel*{
                \begin{tikzpicture}[yscale=-1,transform shape]
                \pic{orcidlogo};
                \end{tikzpicture}
            }{|}}}}

\begin{document}

\title{Development of End-to-End Low-Cost IoT System for Densely Deployed PM Monitoring Network:\\ An Indian Case Study}

\author{Ayu Parmar${\textsuperscript{\orcidicon{0000-0002-6283-6562}}}$, Spanddhana Sara${\textsuperscript{\orcidicon{0000-0001-7965-0521}}}$, Ayush Kumar Dwivedi${\textsuperscript{\orcidicon{0000-0003-2395-6526}}}$,  C. Rajashekar Reddy${\textsuperscript{\orcidicon{0000-0002-9937-5544}}}$, Ishan Patwardhan,\\ 
Sai Dinesh Bijjam, Sachin Chaudhari${\textsuperscript{\orcidicon{0000-0003-2395-6526}}}$, K. S. Rajan${\textsuperscript{\orcidicon{0000-0002-3347-4451}}}$, Kavita Vemuri
\thanks{This work is partially supported by Dept. of Science \& Technology (DST), Govt. of India under grant no. 2073 (2020) and Pernod Ricard India Foundation (PRIF) Social Incubator Program 2019, with no conflict of interest.}
\thanks{Ayu Parmar, Spanddhana Sara, Ayush Kumar Dwivedi, C. Rajashekar Reddy, Ishan Patwardhan, Sai Dinesh Bijjam, Sachin Chaudhari, K. S. Rajan and Kavita Vemuri are with IIIT Hyderabad 500032, India (e-mail: \{ayu.parmar, spanddhana.sara, ayush.dwivedi, rajashekar.reddy, ishan.patwardhan\}@research.iiit.ac.in; sai.dinesh@students.iiit.ac.in; \{sachin.c, rajan, kvemuri\}@iiit.ac.in).}
}
%
%
%
\maketitle
\begin{abstract}
Particulate matter (PM) is considered the primary contributor to air pollution and has severe implications for general health. PM concentration has high spatial variability and thus needs to be monitored locally. Traditional PM monitoring setups are bulky, expensive and cannot be scaled for dense deployments. This paper argues for a densely deployed network of IoT-enabled PM monitoring devices using low-cost sensors. In this work, 49 devices were deployed in a region of the Indian metropolitan city of Hyderabad out-of this, 43 devices were developed as part of this work and 6 devices were taken off the shelf. The low-cost sensors were calibrated for seasonal variations using a precise reference sensor. A thorough analysis of data collected for seven months has been presented to establish the need for dense deployment of PM monitoring devices. Different analyses such as mean, variance, spatial interpolation and correlation have been employed to generate interesting insights about temporal and seasonal variations of PM. In addition, event-driven spatio-temporal analysis is done for PM values to understand the impact of the bursting of firecrackers on the evening of the Diwali festival. A web-based dashboard is designed for real-time data visualization.
\end{abstract}
\begin{IEEEkeywords}
Correlation analysis, Dense deployment, Diwali analysis, Internet-of-things, Particulate matter, Seasonal variation, Spatial interpolation
\end{IEEEkeywords}
%
\section{Introduction}
Air pollution has been an issue of grave concern across the world for decades  \cite{Ayres}. Particulate matter (PM) occurring from local activities is a significant contributor to air pollution, which causes serious health implications. In India, the average PM concentration is 55.8 µg/m³, 11 times higher than the WHO guideline \cite{AQLI}. According to an estimate, in 2019, around 6.7 million premature deaths globally were associated with air pollution \cite{Lancet2022}. Studying and continuously monitoring the various patterns related to air pollution is essential to address the challenge comprehensively. Many countries have established elaborate structures for air-quality monitoring based on beta attenuation monitor (BAM) and tapered element oscillating microbalance (TEOM) often deployed by pollution control boards and other governmental agencies to monitor air quality \cite{Khot}. Although the PM data from these stations is very accurate, this approach has the limitation of scalability. Often the centralized \ac{AQMS} are expensive, bulky and large in size \cite{Ying}. Thus, they cannot be densely deployed. For example, in a big metropolitan city like Hyderabad, with a population of over 6.7 million \cite{population} and area over $650\,\text{km}^2$, the Central Pollution Control Board (CPCB) and the Telangana State Pollution Control Board (TSPCB) have deployed only 12 \ac{AQMS} \cite{CPCB}. All of this points out not only the issue of the expensive and time-consuming setup of the air quality monitoring apparatus but also a  significant mismatch between the requirement of PM data and its availability. Consequently, the resolution of available air quality data is limited as very few stations are typically responsible for an entire city region. Low resolution is not enough for a deeper understanding of PM as the pollutant levels can vary drastically even within smaller blocks in a city \cite{Apte}.

To overcome the limitation of scalability, several studies have been carried out recently focusing on the use of low-cost sensors-based PM monitoring devices \cite{Andrea}. Moreover, internet-of-things (IoT) and cloud computing have been used for real-time PM monitoring. Citywide deployment of such devices has been studied to increase pollution data's spatial-temporal resolution. For instance, in \cite{169China}, 169 devices were deployed in $20\, \text{km} \times 20\,\text{km}$ area measuring PM2.5 every hour with a focus on identifying the primary source of PM2.5. Similarly, in \cite{Denslydeployed100}, 100 devices have been deployed, both stationary and mobile, for 5 months in Turin, Italy. In \cite{50device}, 50 devices have been deployed in a large area of $100\, \text{km}^2$ for 6 months in Utah, United States, discussing the reliability and efficiency of low-cost PM sensors for spatial and temporal dense heterogeneous pollution data. In \cite{30sensor}, 30 devices were deployed in an area of $800\, \text{m} \times 800\,\text{m}$ discussing the suitability of low-cost sensors for networks of air quality monitors for dense deployment, but the data collection is very less. In \cite{45_Popoola}, authors deployed a network of 45 low-cost electrochemical sensor devices in and around Cambridge, UK, for 2.5 months. The network provided measurements of harmful gases (CO, NO, NO2), temperature and relative humidity (RH). The collected data from these devices was analyzed for source attribution. The authors also determined the regional pollution level by studying variations in the sensor readings deployed in different environments. In \cite{Brienza}, authors developed a low-cost device named uSense, which can measure the concentration of harmful gases (NO2, O3, CO) in the ambient environment. It uses Wi-Fi to offload the sensed data on a cloud server. Wi-Fi connectivity enables users to place the device indoors or outdoors in places like balconies or gardens.


For localized monitoring and real-time analysis of outdoor PM, a dense IoT system is needed, which is scalable with low-cost portable ambient sensors. In our preliminary work \cite{rajashekar2020}, a smaller network with 10 devices was deployed inside the IIITH campus for PM monitoring. Field experience from this deployment highlighted the need for a denser deployment across different environments and areas with calibrated PM sensor and a more robust device that can cache data to avoid loss due to communication outages. In this paper, an end-to-end low-cost IoT system is developed and densely deployed for monitoring PM with fine spatio-temporal resolution. The system includes designing the hardware, calibrating the low-cost sensors, cloud interfacing, and developing a web-based dashboard.

The specific contributions of this paper are:
\begin{enumerate}
    \item For the high spatial resolution of outdoor PM, 49 IoT-based PM monitoring devices were developed, calibrated and deployed at various outdoor locations.
    \item The developed device is designed to be robust against the issue of data loss due to connection and power outages. The device maintains an offline cache in the event of an outage. The stored data is offloaded in bulk once the power and communication are restored. 
    \item All PM sensors were calibrated for seasonal variations by co-locating with a reference sensor. Also, each device was calibrated individually.
   \item The devices were deployed at 49 outdoor locations covering a $4\, \text{km}^2$ area in Gachibowli, Hyderabad, India. The field locations were selected to include urban, semi-urban, and green regions. Few devices were deployed at busy traffic junctions and roadsides. The data were recorded at a frequency of every 30 seconds (sec) spanning over all the seasons for six months, thus aggregating 20.7 million data points.
   \item A web-based dashboard was developed and deployed to visualize the data in real-time. 
   \item Different analyses were carried out by observing seasonal mean and variance, spatial interpolation, event-driven variation and correlation. 
   Results show the optimal deployment across a varied landscape and can be a key factor in identifying the release of high concentration in real-time.
\end{enumerate}

Our work differs significantly from the previous studies discussed earlier \cite{169China, Denslydeployed100, 50device, 30sensor, 45_Popoola, Brienza}. For example, \cite{45_Popoola, Brienza} are about gas monitoring, while the main focus of this paper is on PM monitoring. The studies in \cite{169China, Denslydeployed100, 50device} have a large number of devices for citywide monitoring of PM but have a sparse deployment ($<$ 0.6 devices per ${km}^2$) in contrast to our dense deployment ($>$ 12 devices per ${km}^2$). The work in \cite{30sensor} is on dense deployment, but it is not an IoT network. Data is collected locally on microSD card for a small duration, while our work involves an IoT network deployed since last one year collecting large data sets.


The rest of the paper is structured as follows: Section II describes the complete system's hardware architecture and software design. Section III describes the measurement setup and the deployment plan. Section IV  describes the process of data collection, preprocessing and calibration. Section V presents the development of the dashboard and its framework. Section VI presents the observations from the measurement campaign. Finally, the conclusions are deduced and articulated in Section VII.
%
\begin{figure*}[t!]
\centering
\subfloat[Block Architecture]{\includegraphics[width=0.4\linewidth]{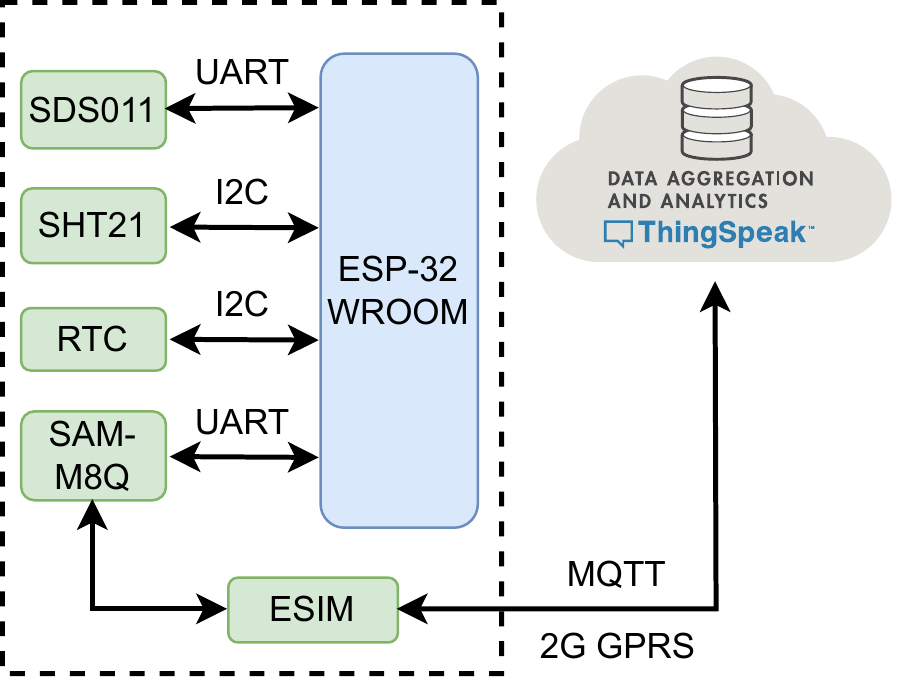}\label{fig:block_arch}}
\hfil
\subfloat[PM monitoring circuit board]{\includegraphics[width=0.53\linewidth]{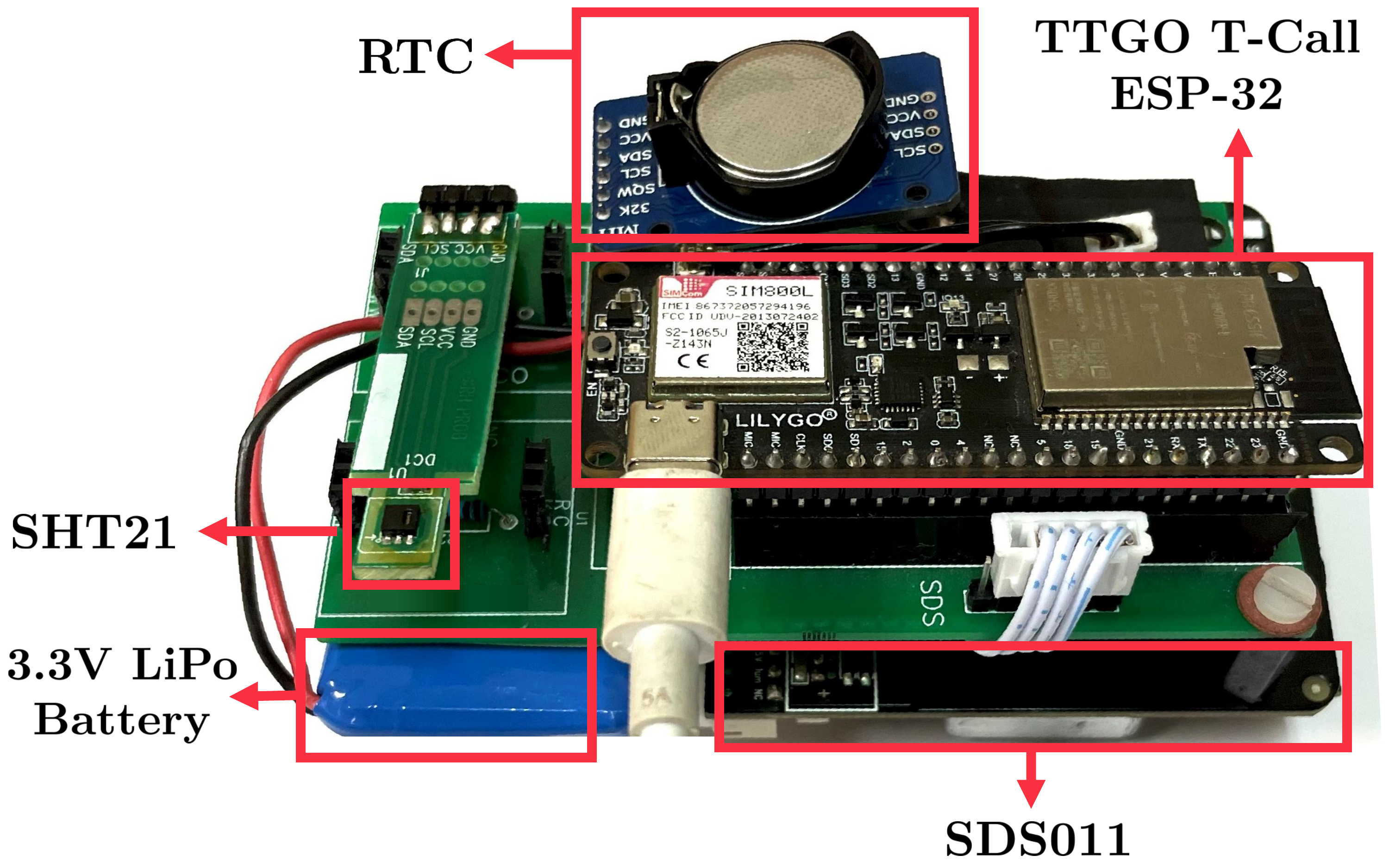}\label{fig:node}}
\caption{Block Architecture and the circuit board of the deployed PM monitoring device.}
\vspace{-0.1cm}
\label{arch_pmnode}
\end{figure*}
%
\section{Hardware Architecture: General Overview And Specifications}
\subsection{Hardware Specification}
%
{\renewcommand{\arraystretch}{1.2}
\begin{table}[t!]
\caption{Specifications of the components used in the developed PM monitoring device.}
\centering
\begin{tabular} {|c|c|c|}
\hline
\bf Component & \bf Specification & \bf Value \\  
\hline \hline
SDS011 \cite{SDS011} & Operating Voltage Range &   4.7 V to 5.3 V\\
& Operating Temperature Range &    $-20\,^\circ$C to $+50\,^\circ$C\\
& Operating Humidity Range &   0 \%RH to 75\,\%RH\\
& Measurement Parameters &     PM2.5 \& PM10\\
& Measurement Particle Size &  0.3 to 10$\,\mu$m\\
& Measuring Range &    0.0 to 999.9\,\si{\micro\gram\per\cubic\meter}\\
& Serial Data Output Frequency &    1\,s\\
& Maximum Current & 100\,mA\\
& Signal Output & UART, PWM\\
\hline
SHT21 \cite{SHT21} & Operating Voltage Range & 2.1 V to 3.6 V\\
& Operating Temperature Range &     $-40\,^\circ$C to $+125\,^\circ$C\\
& Operating Humidity  Range& 0 \%RH to 100 \%RH\\
& Temperature Resolution &    $0.01\,^\circ$C\\
& Humidity Resolution &    0.04\,\%RH\\
& Temperature Accuracy &  $\pm0.3\,^\circ$C \\
& Humidity Accuracy &   $\pm 2.0\, \%$RH \\
& Response Time & 8 s to 30 s\\
& Signal Output & I2C\\ 
\hline
TTGO T-Call & Operating Voltage Range &    3.3 V\\
ESP32 \cite{TTGO}& Operating Temperature Range &  $-40\,^\circ$C to $+85\,^\circ $C\\
& Max Operating Frequency & 240 MHz\\
& RAM & 540 KB\\
& Wi-Fi& IEEE 802.11 b/g/n\\
& SIM Module & SIM800L\\
\hline
eSIM \cite{E-sim} & Operating Voltage Range &   1.62 V to 5 V\\
& Operating Temperature & $-40\,^\circ$C to $+105\,^\circ$C\\
& Available Memory & 128 KB or more\\
& Technology & 2G GPRS\\
& Bandwidth & 25 MHz\\
\hline
\end{tabular}
\label{SensorSpecifications}
\end{table}}
%
%
Fig. \ref{arch_pmnode} shows the hardware architecture and circuit board for the developed PM monitoring device in this work. The basic architecture consists of sensors (PM sensor SDS011 and temperature humidity sensor SHT21), a communication module (SIM800L and eSIM), a real-time clock (RTC), and a lithium polymer (LiPo) battery. 
All these components are connected to the microcontroller TTGO T-Call ESP32. The controller reads data from all the sensors periodically every 30 sec and offloads it to ThingSpeak, a cloud-based server employing message queuing telemetry transport secured (MQTTS) over a 2G or Wi-Fi network and the data packet size is 28 bytes. The device is powered with an AC-DC power adapter and a 1000 mAh battery and enclosed in an IP65 box made of ABS filament, the enclosure offer complete protection against dust and a good level of protection against water. The form dimensions are: width = 125 mm, depth = 125 mm and height = 125 mm.  The SDS011 and SIM800L modules are connected to the controller through the UART protocol, while the SHT21 and RTC are connected through the I2C protocol. The overall cost of the device after adding the cost of individual hardware components is \rupee 7000 (approximately 85 USD)\footnote{Assuming the conversion rate of 1 USD = \rupee 82.35 in October 2022.}. The specifications of the individual hardware components are listed in Table \ref{SensorSpecifications}. The details of each component are given below.
\subsubsection{Nova SDS011}
A Nova-SDS011 PM sensor \cite{SDS011} is used in this study to gauge the presence and concentration of tiny particles with a diameter \SI{10}{\micro\metre} or less (referred to as PM10) and particles with a diameter \SI{2.5}{\micro\metre} or less (referred as PM2.5). The sensor uses light scattering phenomena to sense particle concentration. 
It is a low-cost sensor with a good correlation with BAM for PM monitoring \cite{Badura2018}. At the same time, it can achieve low error values after calibrating and using a simple linear regression \cite{patwardhan2021}.  


%
\subsubsection{SHT21} 
At high temperature and RH levels, light scattering-based PM sensors do not operate reliably \cite{humiditydependent}. Therefore, temperature-humidity sensor SHT21 \cite{SHT21} is installed in the developed device to measure temperature and RH for filtering out any unreliable PM values.
\subsubsection{Battery/Power Adapter}
The developed device is powered using a 3.3 V rechargeable 1000 mAh LiPo battery. A battery management circuit onboard the micro-controller module and a dedicated AC-to-DC power adapter are used to charge the battery. 
If the AC input is available, the device will work using an AC-to-DC power adapter and charge the battery. It will switch automatically to battery power if no AC input is available.
%
\subsubsection{TTGO T-Call ESP32}
In field deployment, a wireless module is needed to send the sensed data to the cloud. This study uses the TTGO T-Call ESP-32 \cite{TTGO}, a Wi-Fi and SIM800L GSM/GPRS module with built-in Bluetooth wireless capabilities. Each device is configured to use Wi-Fi or GPRS according to the network availability in the area.
\subsubsection{Cellular Network}
In the previous deployment on the campus, \cite{rajashekar2020}, Wi-Fi routers were used to communicate sensed data to the ThingSpeak server. However, Wi-Fi connectivity is unavailable at most locations outside the campus, particularly on the roadsides. In such places, we have used a 2G network, which has very good coverage in the city. Since the sensed data is small in packet size (24 bytes), a 2G network data rate is sufficient. Moreover, the use of removable SIM cards in-field deployment gives rise to the threat of their misuse in case of theft. Hence an embedded SIM (eSIM) customized for IoT applications from the Sensorise service provider is used in the device \cite{E-sim}. An eSIM is a small chip embedded in the hardware setup with the help of the SIM 800L module\cite{eSIM}, thus restricting the reusability by the general public. IoT devices planned for long-term projects and having a field deployment are protected from the impact of evolving network technologies or service terminations by eliminating technical or carrier lock-ins with a single eSIM. The information on eSIM is rewritable, which makes it easy to change the operator at any time. It allows the user to change the service provider over the air without physically changing the eSIM. The Sensorise eSIM also provides the facility of multi-operator subscriptions. If there is an issue with one network, eSIMs can switch to other operators to connect to the network.
\subsubsection{ThingSpeak}
In the deployed IoT network, the data is aggregated at ThingSpeak, a cloud-based platform for IoT applications. It facilitates data access, logging and retrieval by providing an application programming interface (API) \cite{Thingspeak}. In \cite{Ihitasecurity}, the authors conducted a security analysis of AirIoT, an air-quality monitoring network and proposed solutions for baseline security of any smart city air monitoring network. The paper recommended using MQTTS instead of MQTT as the latter does not provide data integrity, while attackers can access information such as payload, topic names, and IP addresses. Therefore, the MQTTS protocol is implemented in this work for communication between the device and the ThingSpeak.
%
\begin{figure}[!t]
\centering
\includegraphics[width=0.902\linewidth]{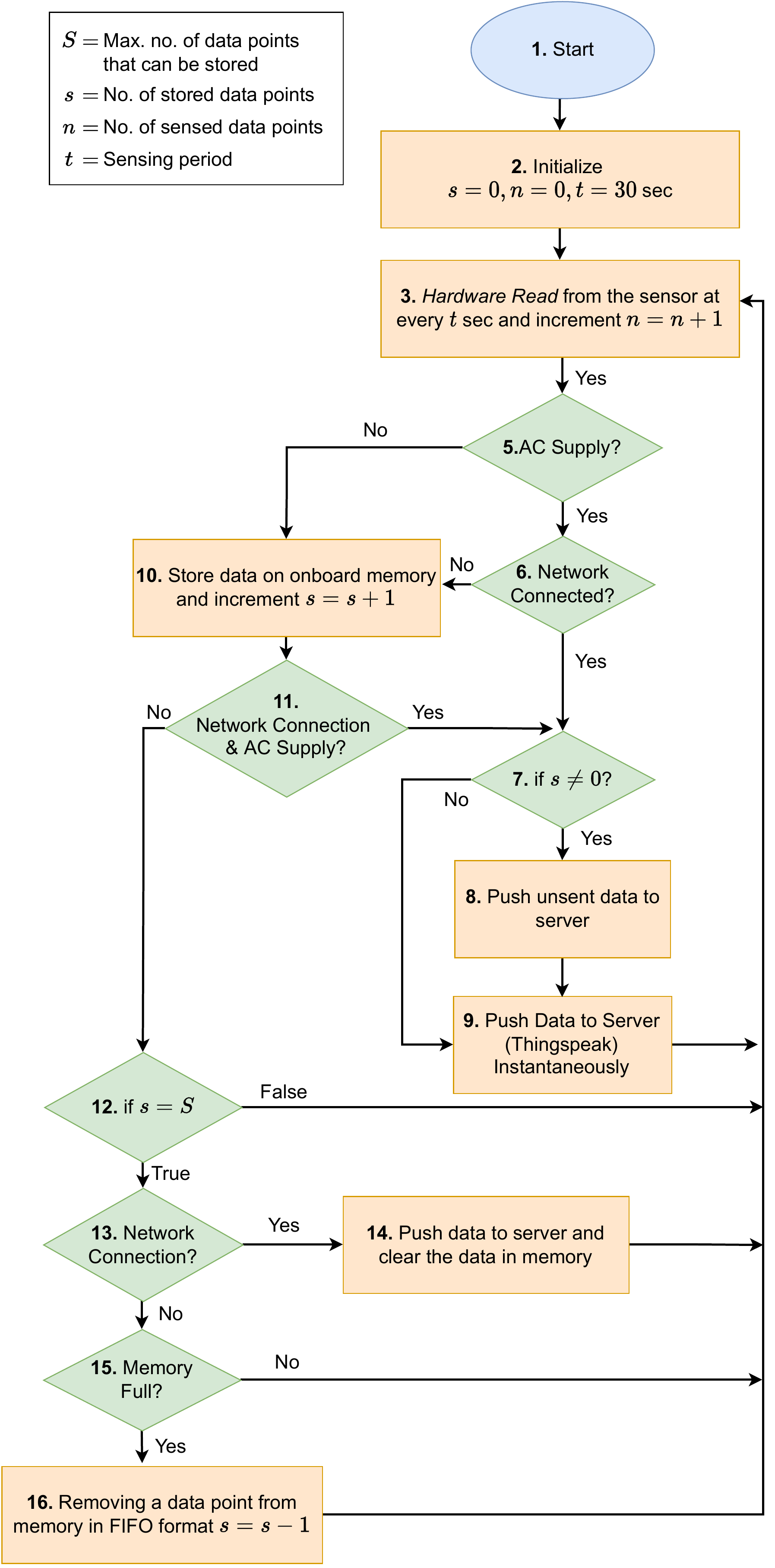}
\caption{\centering Working mechanism of the device.}
\label{fig:flow diag}
\end{figure}
%
%
\subsection{Working mechanism of the device}\label{flowChart}
Fig.\ref{fig:flow diag} illustrates the flowchart of the sensing algorithm developed to avoid data loss in the event of a connection outage. The microcontroller first reads the sensed data every 30 sec; however, the time to offload the sensed data depends upon the network. Next, the controller checks the network connectivity for pushing the data to the server. If the network is available, the data is transmitted instantaneously. However, if the network is unavailable, the data is stored locally in a part of the microcontroller RAM until the device reconnects to the network. Note that the size of microcontroller RAM  is 540 KB and part of it is used for code and header files (created while pushing data), while the part of the remaining memory can be used to store data. We define $S$ = 20000 as the maximum number of data points that can be stored. 
Once the connection restores, the stored data is uploaded to the cloud server in a bulk transmission and subsequently cleared from the device. In the case device memory is filled with back-logged data in the event of a long connection outage, i.e., if the number of stored data points ($s$) is equal to the $S$, the data is cleared in first-in-first-out (FIFO) format to make space for the new incoming data.
%
\begin{figure*}[t!]
\centering
\subfloat[Deployment plan]{\includegraphics[width=0.26\linewidth]{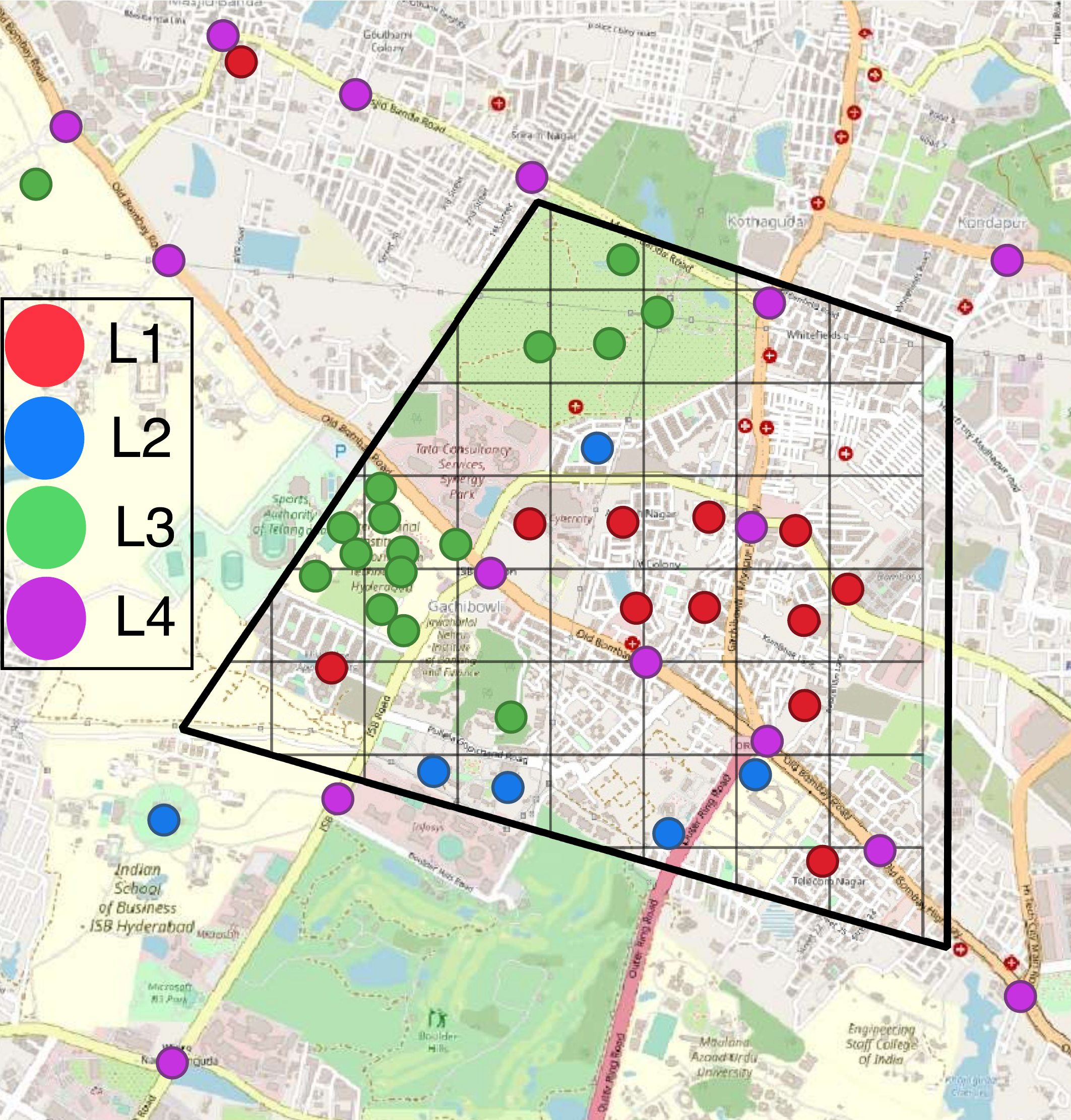}\label{fig:Node location}}
\hfil
\subfloat[Example field deployment]{\includegraphics[width=0.27\linewidth]{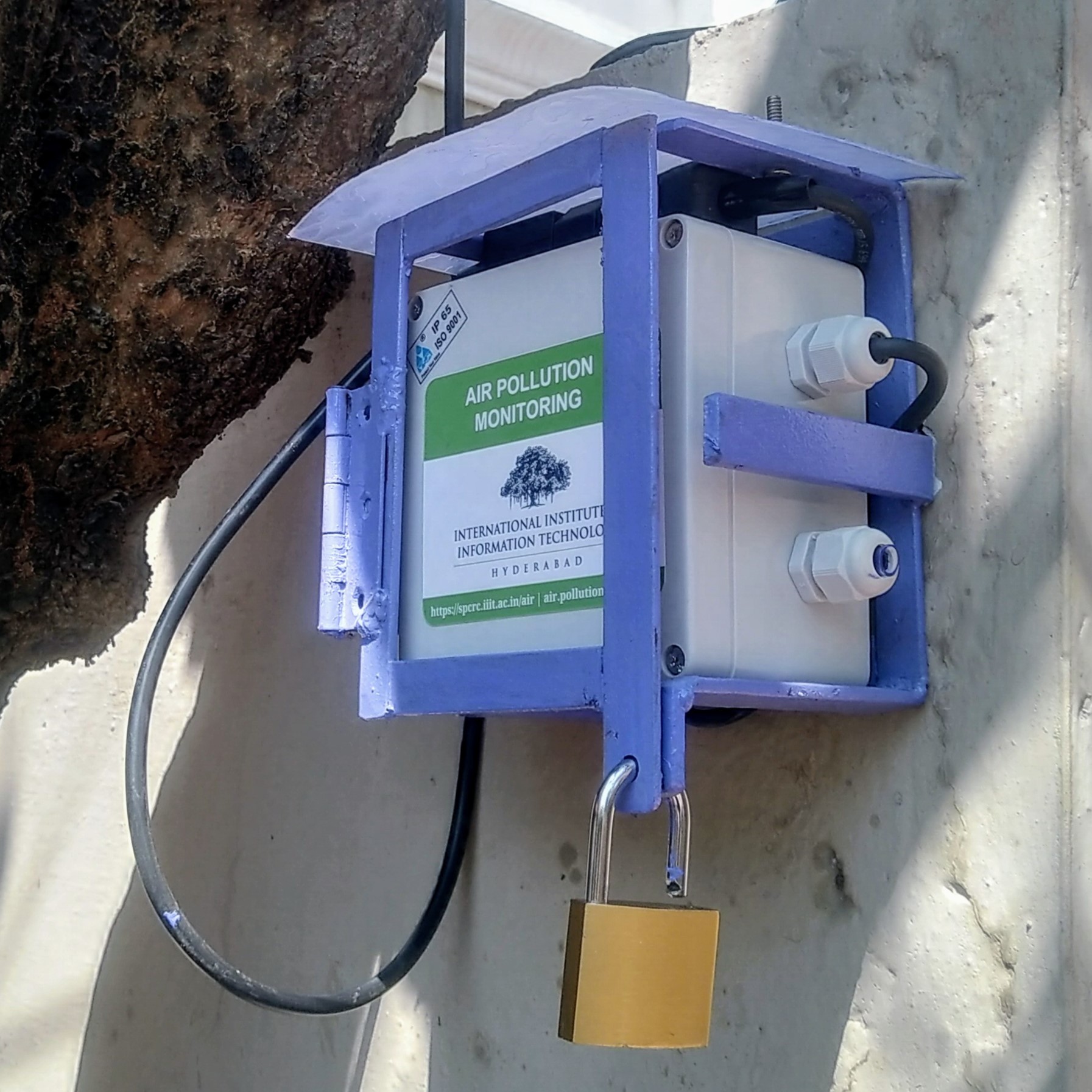}\label{fig:deployment}}
\caption{Deployment plan covering urban, semi-urban, green region, junctions, and roadsides poles.}
\vspace{0.1cm}
\end{figure*}
%
\begin{figure*}[t!]
\centering
\includegraphics[width=0.7\linewidth]{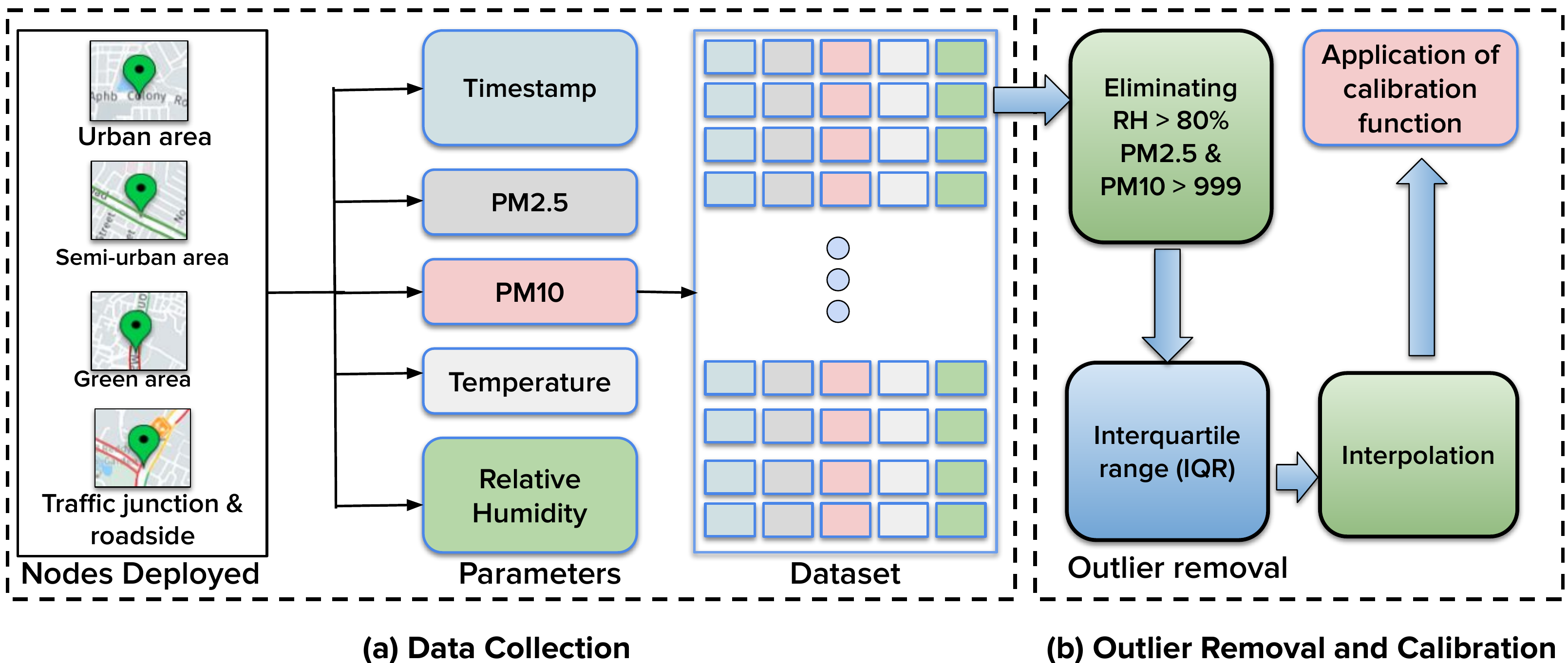}
\caption{Data collection, preprocessing, and calibration.}
\label{fig:datapreprocessing}
\end{figure*}

\section{Deployment Strategy}
Fig. \ref{fig:Node location} shows the plan for the field deployment of devices, while Fig.\ref{fig:deployment} shows an example of the deployed device at one of the locations. 
The deployment was done in the Gachibowli region of Hyderabad, the capital city of the Telangana state and the fourth largest populated city in India 
\cite{population}. A total of 49 devices were deployed in a region of approximately $4\,\text{km}^2$ to understand the variation of PM across different environments and areas.

%
\begin{figure*}[t!]
\centering
\subfloat[Timeseries Plot]{\includegraphics[width=0.505\linewidth,trim={0.1cm 0.2cm 0.2cm 0.7cm},clip]{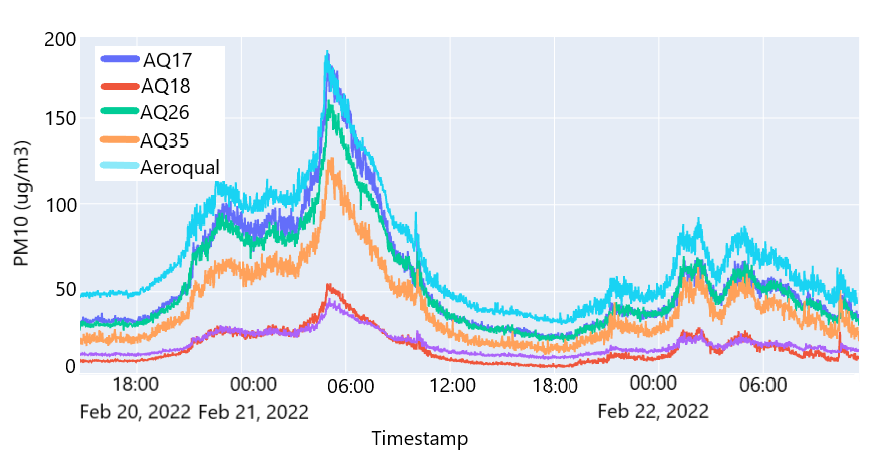}
\label{fig:PM10_raw1}}
\hfil
\subfloat[Scatter Plot]{\includegraphics[width=0.25\linewidth]{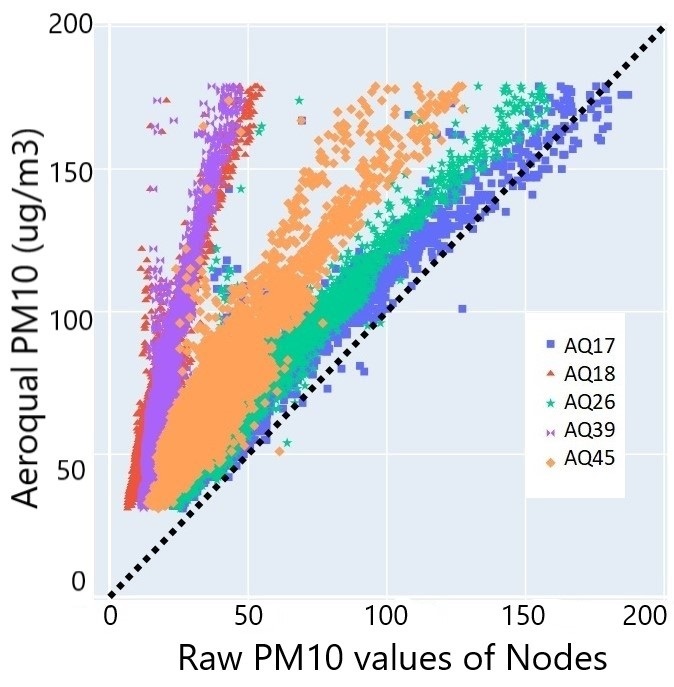}\label{fig:scatter plot}}
\caption{Time series and scatter plot of raw PM10 data (1-hour average).}
\label{fig:rawOverall}
\end{figure*}
%
\begin{figure*}[t!]
\centering
\subfloat[Timeseries Plot]{\includegraphics[width=0.505\linewidth,trim={0.1cm 0.2cm 0.2cm 0.6cm},clip]{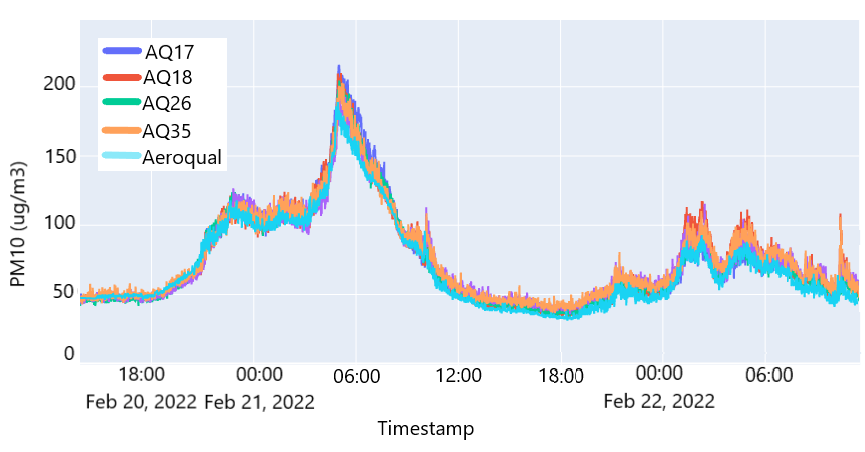}\label{fig:PM10_calibrated}}
\hfil
\subfloat[Scatter Plot]{\includegraphics[width=0.25\linewidth]{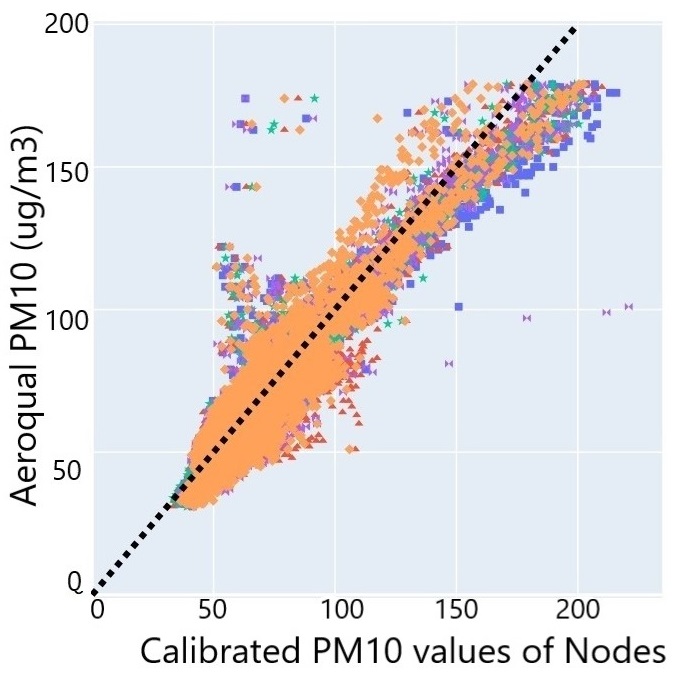}\label{fig:Calibrated_Data}}
\caption{Time series and scatter plot of calibrated PM10 data (1-hour average).}
\label{fig:calibratedOverall}
\end{figure*}
%
%
\begin{figure*}[t!]
\centering
\subfloat[Default view of the dashboard]{\includegraphics[width=0.9\linewidth]{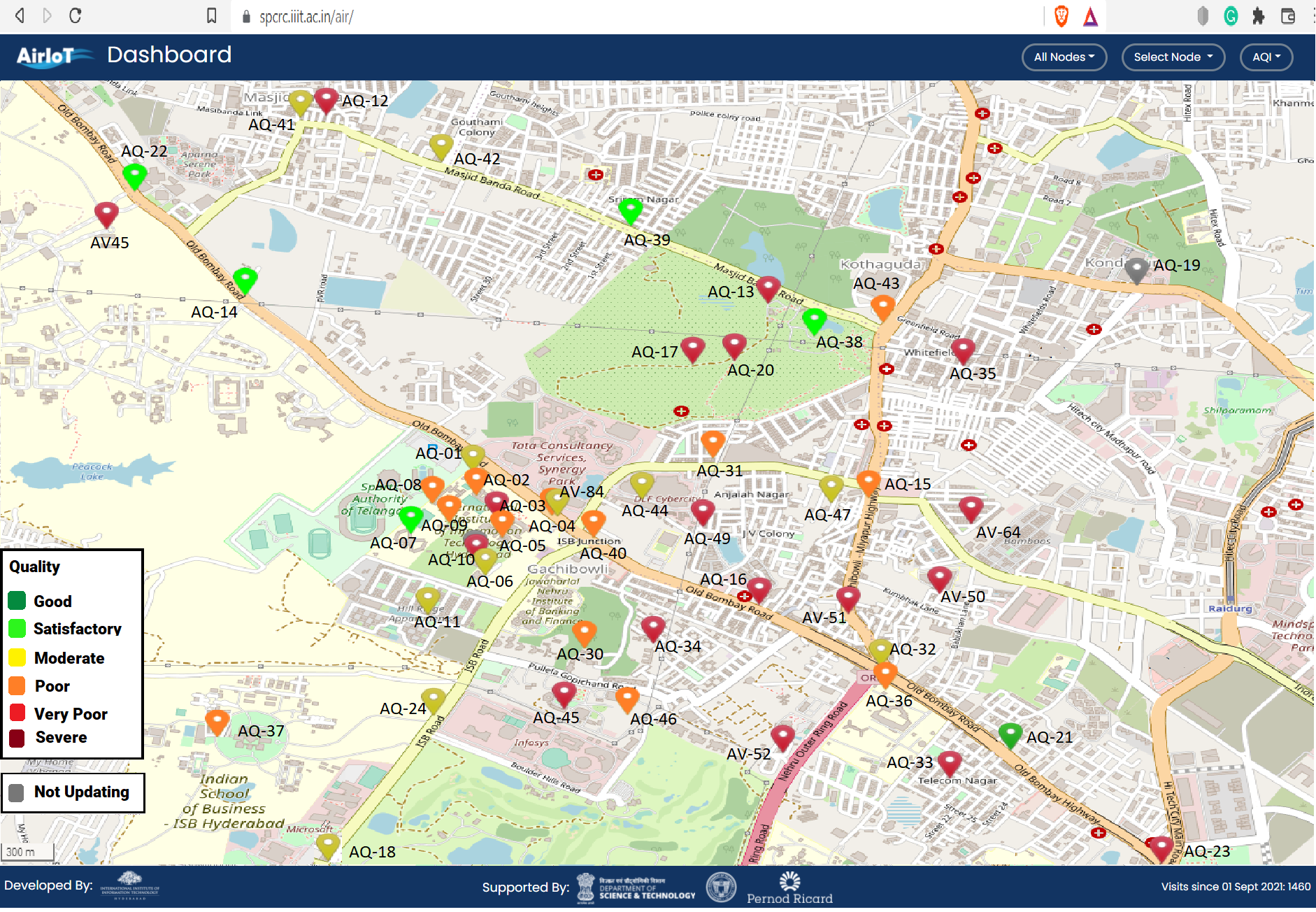}\label{fig:dashboard}}
\vspace{5mm}
\subfloat[Graphical representation of the real-time data]{\includegraphics[width=0.46\linewidth]{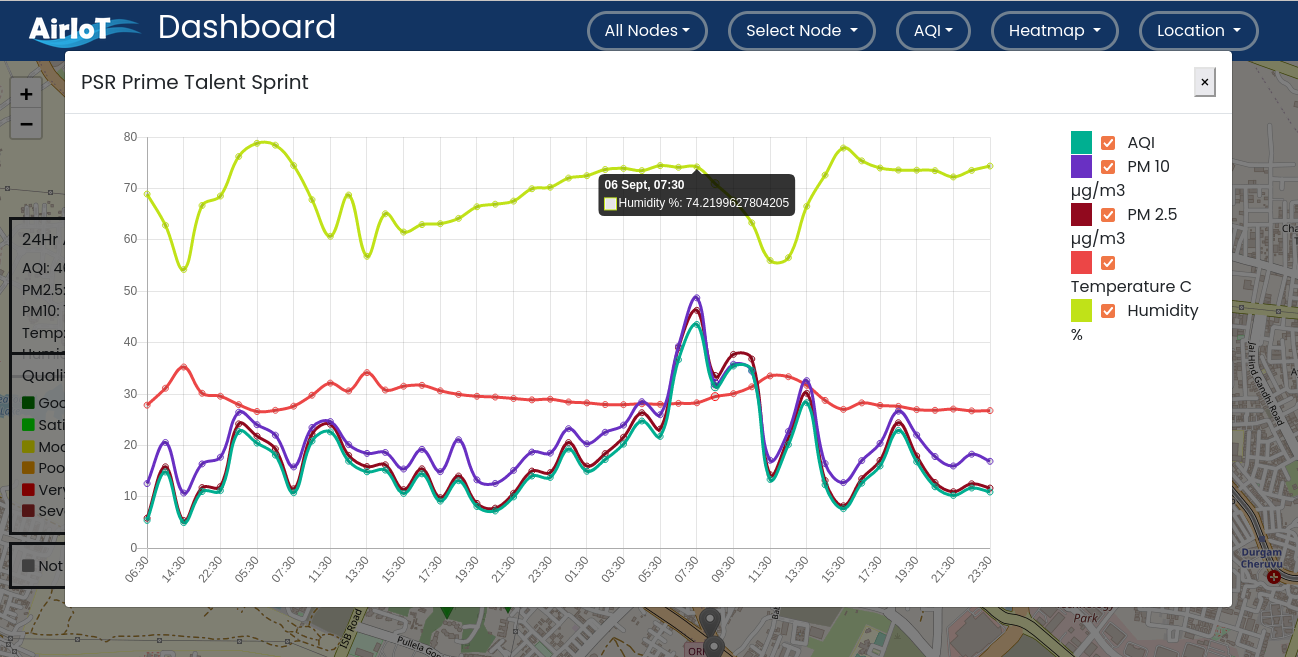}\label{fig:dashboardgraph}}
\hfill
\subfloat[Architecture of the dashboard]{\includegraphics[width=0.45\linewidth]{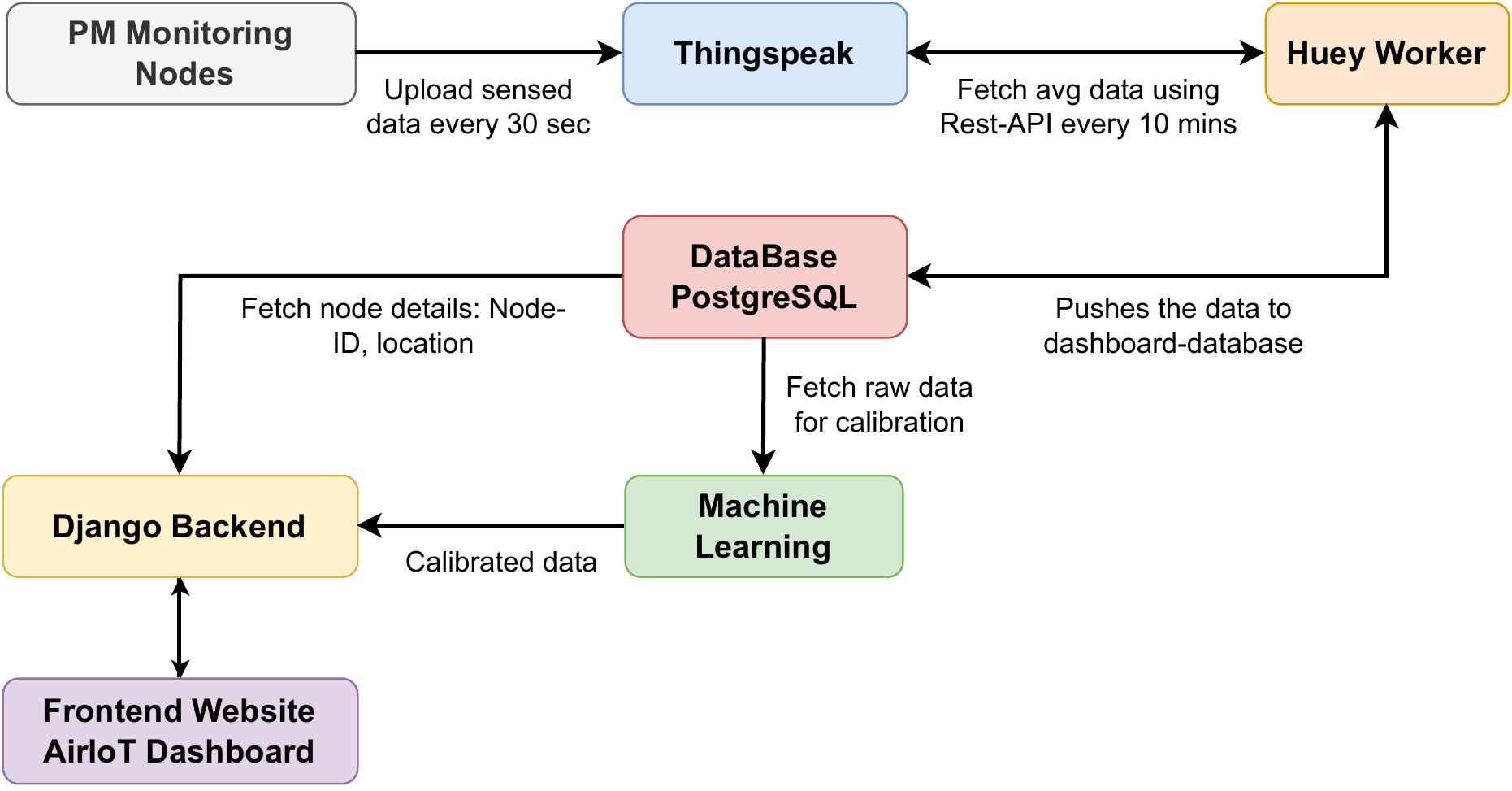}\label{fig:dashboardArch}}
\vspace{0.5cm}
\subfloat[Data-flow pipeline of the dashboard]{\includegraphics[width=0.5\linewidth]{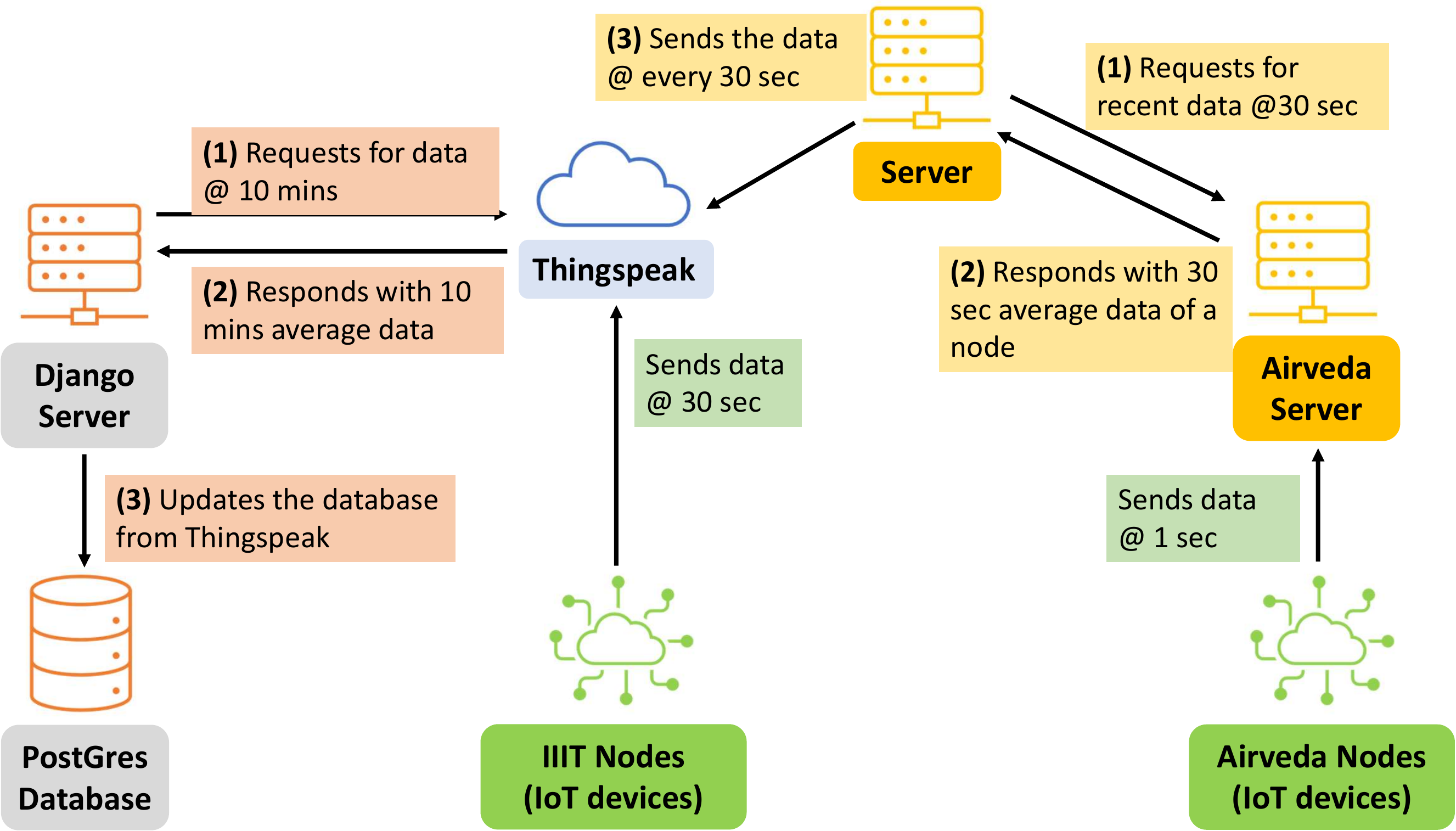}\label{fig:dashboardBackend}}
\hfill
\subfloat[Communication between the users and the server]{\includegraphics[width=0.45\linewidth]{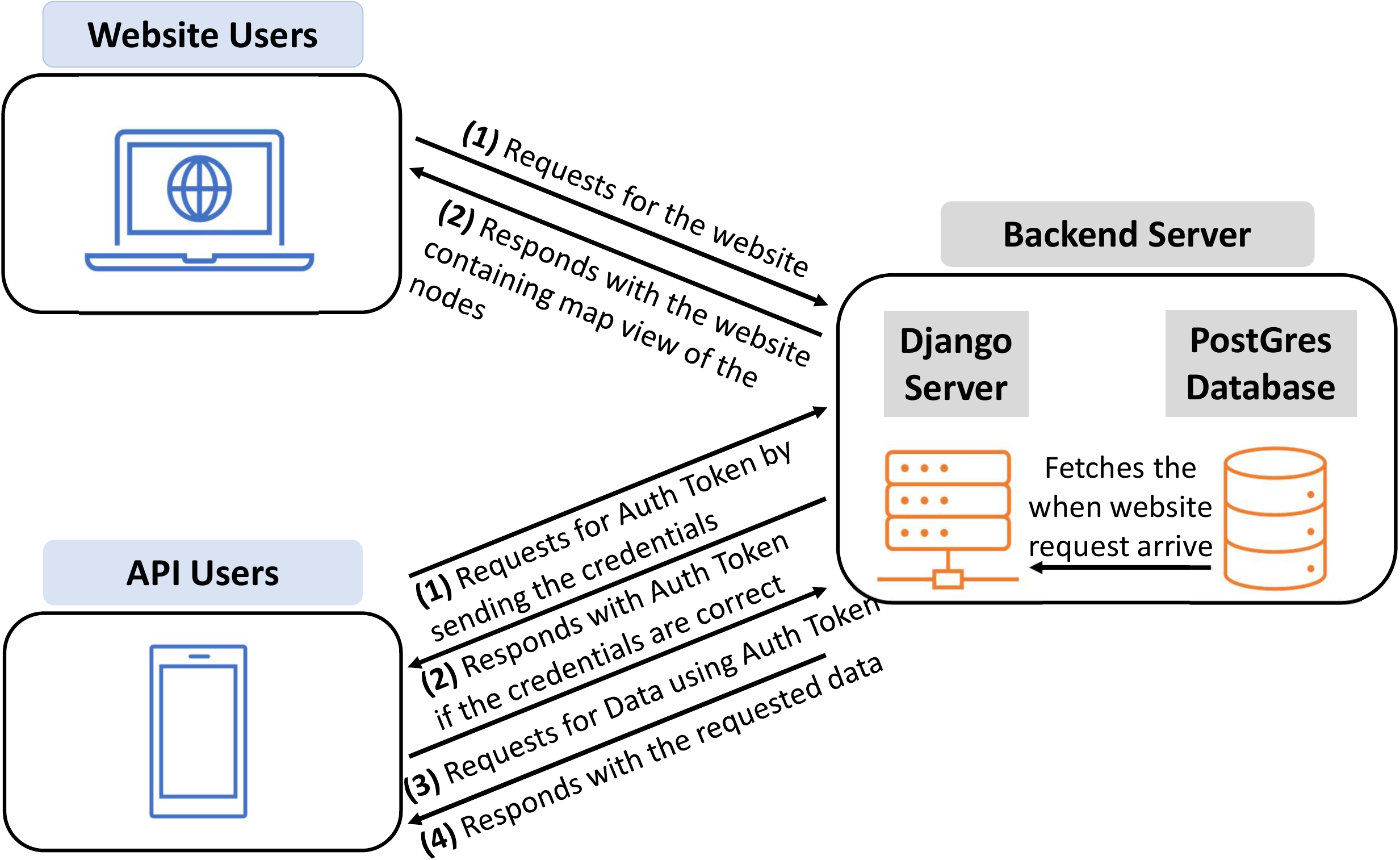}\label{fig:dashbaordFetch}}
\caption{Web-based dashboard for real-time visualization of air quality data from all the deployed devices.}
\label{fig:dashboardOverall}
\end{figure*}
%
%
Based on the landscape pattern of Gachibowli, the entire region is divided into three categories: urban, semi-urban and green. A few devices have also been deployed at busy traffic junctions and roadsides. Fig. \ref{fig:Node location} shows the deployment plan of all the devices with exact locations of the following location types:
\begin{itemize}
    \item Location type L1:  Urban region
    \item Location type L2: Semi-urban region
    \item Location type L3: Green region
    \item Location type L4: Traffic junctions and roadsides poles
\end{itemize}
%
%

The $4\,\text{km}^2$ area has been divided into approximately 42 boxes. Every square box in Fig. \ref{fig:Node location} represents an area of $400\times400 \text{ m}^2$. An attempt has been made to deploy 1 device in each box depending on the availability of power, network and consent availability. However, in some boxes, more than 1 device has been deployed, as shown in \ref{fig:Node location}. 
The field deployment of devices was completed in July 2021 and the data collection started in Aug. 2021. As part of experimentation, along with the devices developed at IIITH, a few devices from an Indian manufacturer, Airveda, were also deployed \cite{Airveda}. Table \ref{DeploymentSetup} summarizes all the deployed devices with their network configuration.
%
{\renewcommand{\arraystretch}{1.29}
\begin{table}[t!]
\caption{Deployment Setup}\label{DeploymentSetup}
\centering
\begin{tabular} {|c|c|c|c|}
\hline
\bf Device & \bf No. of &  \bf Location Type & \bf Network Type \\
& \bf Location & & \\
\hline\hline
IIITH (AQ) &  43 & L1 (07 devices) & Wi-Fi (2 devices)\\
&  & L2 (05 devices) & 2G eSIM (32 devices)\\
&  & L3 (15 devices) & 4G Jio-Fi (9 devices)\\
&  & L4 (16 devices) & \\
\hline
Airveda (AV) & 6 &L1 (04 devices) & Wi-Fi (4 devices)\\
&  & L2 (01 device)& 2G eSIM (2 devices)\\
&  & L3 (01 device) & \\
\hline
\end{tabular}
\end{table}}
%
\section{Data Collection, Preprocessing and Calibration}
Fig. \ref{fig:datapreprocessing} shows a flow diagram depicting different steps in collecting usable data analysis data. This involves data collection, creating the dataset, removing outliers and interpolating missing data, followed by calibration. Each of the steps involved is explained in detail.
\subsection{Data collection}
To create the data set, the air quality was sensed at a frequency of $t$ = 30 \si{\sec} for 43 IIITH devices. For 6 Airveda devices, $t$ = 1 \si{\sec}, averaged over 30 sec.
All the devices were deployed for almost one year and are still deployed. However, usable data were collected for seven months (Aug. 2021, Nov. 2021, Dec. 2021, Jan. 2022, Apr. 2022, May 2022, and June 2022). The loss in the data is because the devices had to be brought back to the lab due to the frequent failure of low-cost sensors requiring regular repair and maintenance. Additionally, the devices were brought for seasonal calibration at regular intervals and to make a major upgrade in the use of ThingSpeak from MQTT to MQTTS (in Mar. 2022). A total of 20.70 million usable data points have been collected. As shown in Fig. \ref{fig:datapreprocessing}a, the collected dataset has PM2.5, PM10, temperature and RH parameters. All the concentration values of PM10 and PM2.5 are mentioned in \si{\micro\gram\per\cubic\meter} hereafter. The temperature and RH values are mentioned in \si{^\circ C} and \si{\%}, respectively. Corresponding to every device, a vector of data points sent from the device is stored on the cloud server for each sensing instance having the following elements:
\begin{itemize}
    \item \textbf{created\_at}: Timestamp at which the sensor value is read. This timestamp is recorded utilizing the RTC module of the device.
    \item \textbf{PM10}: Raw concentration of PM10 read by SDS011.
    \item \textbf{PM2.5}: Raw concentration of PM2.5 read by SDS011.
    \item \textbf{RH}: Raw RH value read by SHT21
    \item \textbf{Temp}:  Raw temperature value read by SHT21 sensor.
    
\end{itemize}
The size of this payload (or sensor data sent from the device) for each sensing instance is 24 bytes. In addition to this, the following static information is  stored in the cloud server 
\begin{itemize}
    \item \textbf{Device\_id}: ID for device identification like IIITH device as AQ-XX and Airveda device as AV-XX, where XX denotes the device sequence number.
    \item \textbf{Location}: Latitude and longitude according to the deployment location.
\end{itemize}
%
\subsection{Data preprocessing}
The following methods have been employed for preprocessing the raw data received from the PM monitoring device:

  \subsubsection{Outlier removal} Environmental conditions like RH and temperature, sensor behavior and anthropogenic activities occasionally result in outliers in the sensed data. Hence the raw data received from the devices need to be preprocessed to make it statistically significant, as shown in Fig. \ref{fig:datapreprocessing}b. PM values are unreliable at higher RH levels (RH$>$80\%). Apart from this, errors may cause raw values to be out of the PM sensor range (0-999). These unreliable points are thus removed. In the dataset, nearly 0.5\% values have been found unreliable. 
  
Further, to identify and remove outliers, the interquartile range (IQR) method \cite{IQR} is used. In this method, the data is separated into four equal parts and sorted in ascending order using three quartiles ($Q_1$, $Q_2$ (median), $Q_3$). Let the difference between the first ($Q_1$) and the third quartile ($Q_3$) be represented by the $I_{QR}$, which is a measure of dispersion. A decision range is set to detect outliers with this approach, and every data point that falls outside this range is deemed an outlier. The lower and upper values in the range are given by
\begin{eqnarray}\label{IQR_method_U}
L_r &=& Q_1 - 1.5 \ I_{QR},\\
U_r &=& Q_3 + 1.5 \ I_{QR}.
\end{eqnarray} 
Any data point less than the $L_r$ or more than the $U_r$ is called an outlier. In the collected dataset, nearly 1.4\% values have been found as an outlier.

\subsubsection{Interpolation} 
Interpolation is a technique to estimate the missing (or removed) data point between two existing data points. In the data set, only 1.9\% data is an outlier which is very less and easy to interpolate. In this work, simple linear interpolation was used for this purpose.

\subsection{Calibration}
For calibration, the low-cost PM sensors were co-located with a reference sensor (Aeroqual S500 \cite{Aeroquals500, AeroqualsPM}) in a ventilated room for a week. Data points were collected at a frequency of 30 sec. A raw dataset of approximately 20,160 data points for each sensor was collected to perform the calibration. Fig. \ref{fig:PM10_raw1} shows the time series plot of PM10 averaged hourly for a few devices before deployment in the field. It can be observed that all the sensors follow the reference sensor in trend but differ with an offset in absolute value. Therefore, there is a need for calibration. It can also be observed that the offsets for each sensor are different. Although not shown to maintain brevity, the same is also valid for raw PM2.5 values.

This paper uses simple linear regression to compensate for the difference between the values of the low-cost sensor and the reference sensor. Linear regression has been chosen since it can compensate for the offset well while preserving the trend in the data, as shown for SDS011 in \cite{patwardhan2021}. The calibrated data $y(i)$ corresponding to the $i^{\text{th}}$ data point can be written as
\begin{equation}
y(i) = m\,x(i)+c,
\label{linearregression}
\end{equation}
where $x(i)$ is the $i^{\text{th}}$ raw data point and $m, c$ are the learned parameters. Each sensor will have a different value of $m$ and $c$.

Fig. \ref{fig:PM10_calibrated} shows the calibrated data of PM10 for a few devices. It can be observed that the low-cost sensors match well with the reference sensor after calibration. Similar results are obtained by training separate functions for PM2.5 as well. It can also be observed from Fig. \ref{fig:scatter plot} and Fig. \ref{fig:Calibrated_Data} that every low-cost sensor differs uniquely from the reference sensor. The same is also concluded in \cite{patwardhan2021} for 3 sensors. Therefore, PM10 and PM2.5 of each low-cost sensor have to be calibrated using unique functions for each sensor. Moreover, it is observed that the sensor behaves differently in different seasons. Hence separate calibration functions have been calculated for different seasons by repeating the process at the season's onset.
%
%
%
%

\begin{figure*}[t!]
\centering
\subfloat[PM10 mean]{\includegraphics[width=0.5\linewidth]{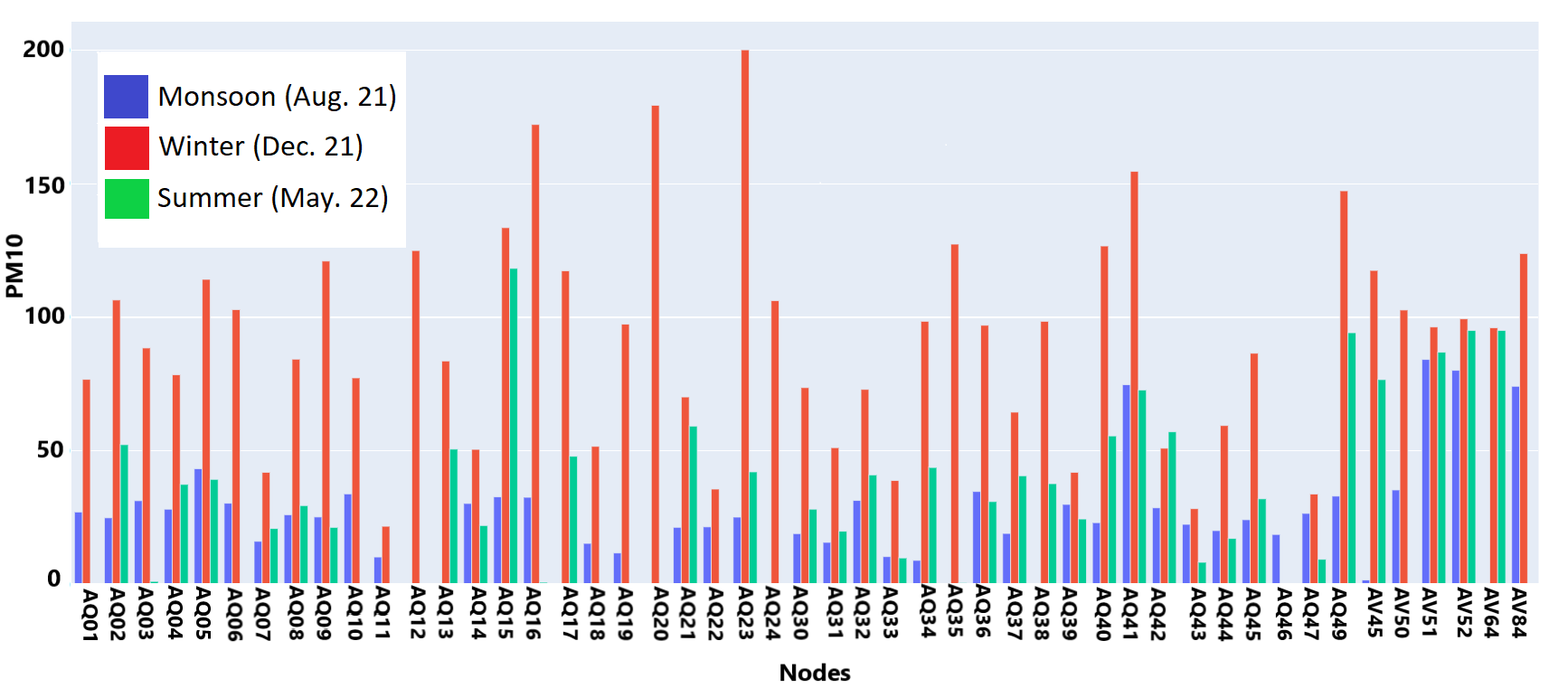}\label{mean10}}
\hfil
\subfloat[PM10 variance]{\includegraphics[width=0.5\linewidth]{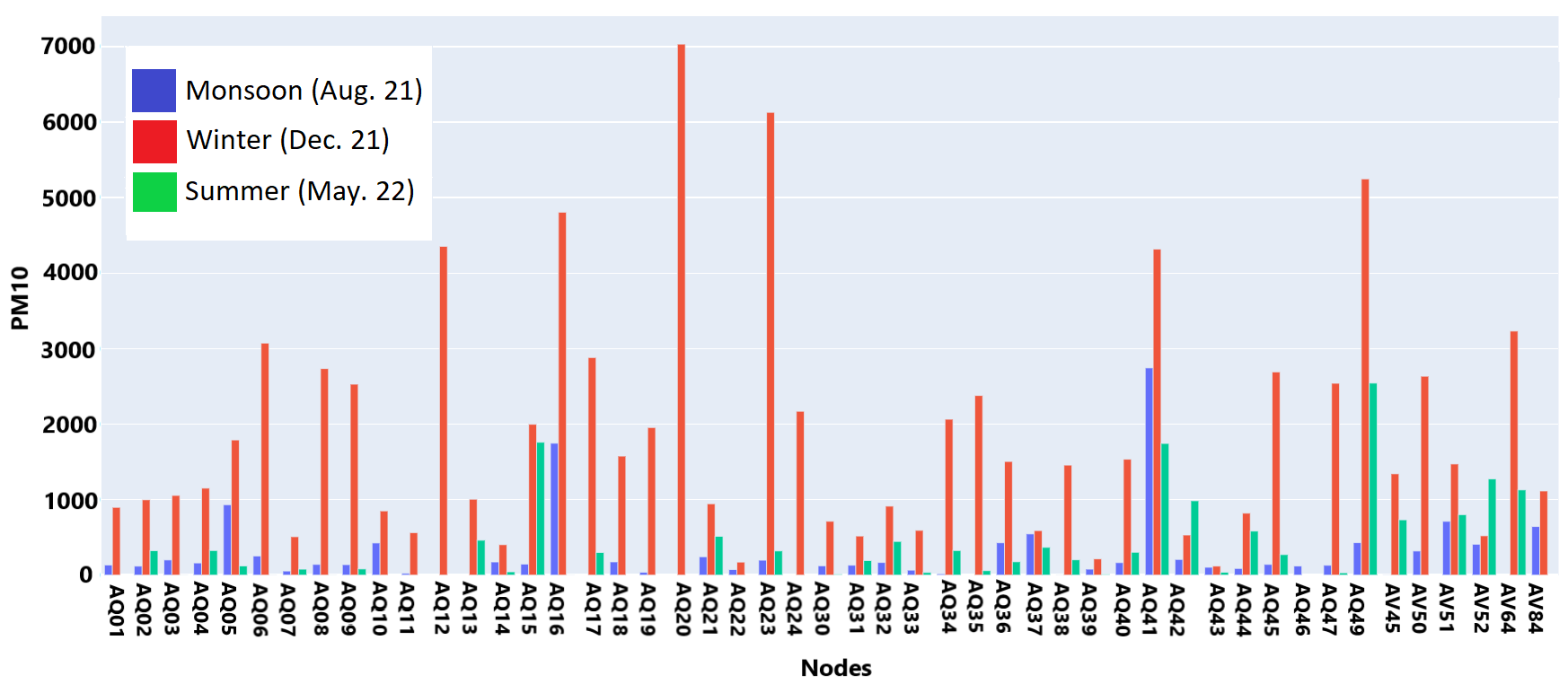}\label{var10}}
\caption{Mean and variance of PM10 concentration at the different locations in different seasons.}
\label{mean_var_pm10}
\end{figure*}
%

\section{Development of web-based dashboard}
Fig. 7 shows the overview of a web-based dashboard named AirIoT, developed to easily access the real-time data sensed by the deployed sensor devices. AirIoT is a responsive website (https://spcrc.iiit.ac.in/air) that aims to visualize the air pollution levels of the areas where the devices have been deployed. Fig. \ref{fig:dashboard} shows the dashboard that provides a view of all the deployed devices and their locations on an interactive map. Fig. \ref{fig:dashboardgraph} shows that for all locations, the users can get 10 minutes average of real-time data and also visualize the trend of the past 48 hours. The user can select and deselect the parameters to visualize the trend individually. The dashboard also provides the features of implementing calibration algorithms, authorizing and maintaining user profiles and logging network activities. Fig. \ref{fig:dashboardArch} shows the basic architecture of the dashboard connecting all the building blocks. The data flow pipe and the sequence of communication steps between the users and the server are presented in Fig. \ref{fig:dashboardBackend} and Fig. \ref{fig:dashbaordFetch}, respectively.

Fig. \ref{fig:dashboardArch} shows the backend of the dashboard, which is built using \textit{Django}. It is an open-source python-based framework created using model-template-views architecture. It provides a host of features that efficiently help the data manipulation using GUI. \textit{PostgreSQL}, an open-source and SQL-compliant relational database management system, is used to store the device's data on the server \cite{PostgreSQL}. The periodic task of fetching data from the ThingSpeak cloud server to the dashboard is implemented using \textit{Huey}, a python queue. Web technologies like CSS, Bootstrap and Javascript are used to build the dashboard's frontend.

The 43 IIIT devices upload the sensed data on the ThingSpeak server every 30 sec, and the 6 Airveda devices upload sensed data on the Airveda server every 1 sec. As shown in Fig. \ref{fig:dashboardBackend}, for data management on a single platform, all the data from the Airveda server is aggregated at ThingSpeak using a separate REST-API with periodic calls. The dashboard fetches data from ThingSpeak for all the locations every 10 minutes and updates the AQI in real-time.

The Django frameworks also help maintain users and their access rights. Fig. \ref{fig:dashbaordFetch} shows that there are mainly two types of supported users, the website users and the API users. The website users are the system administrators who log in on the admin console to maintain or update network details like adding new devices or updating information like location, sensing parameters, calibration coefficients, etc. On the other hand, the API users are other platforms like mobile applications which can take air quality data from the dashboard to integrate with their services. Every API user must use OAuth authentication to fetch the data from the server. Initially, the API users send their credentials to the server. The server verifies them and responds with a token. This token is used in every request to the API to fetch the server's data after that. 
%

\section{Results and Analysis}
This section presents mean and variance analysis results and the spatial interpolation for PM values in different seasons. Further, the event-driven variation analysis is done for the data collected during the festival of Diwali. This is followed by correlation analysis to understand the range, after which the correlation between the two points is insignificant. Note that the results are shown only in terms of PM10 for brevity and similar observations have been made for PM2.5 as well.  

%
\begin{figure*}[tbh]
\centering
\subfloat[At 1100 hrs]{\includegraphics[width=0.3\linewidth]{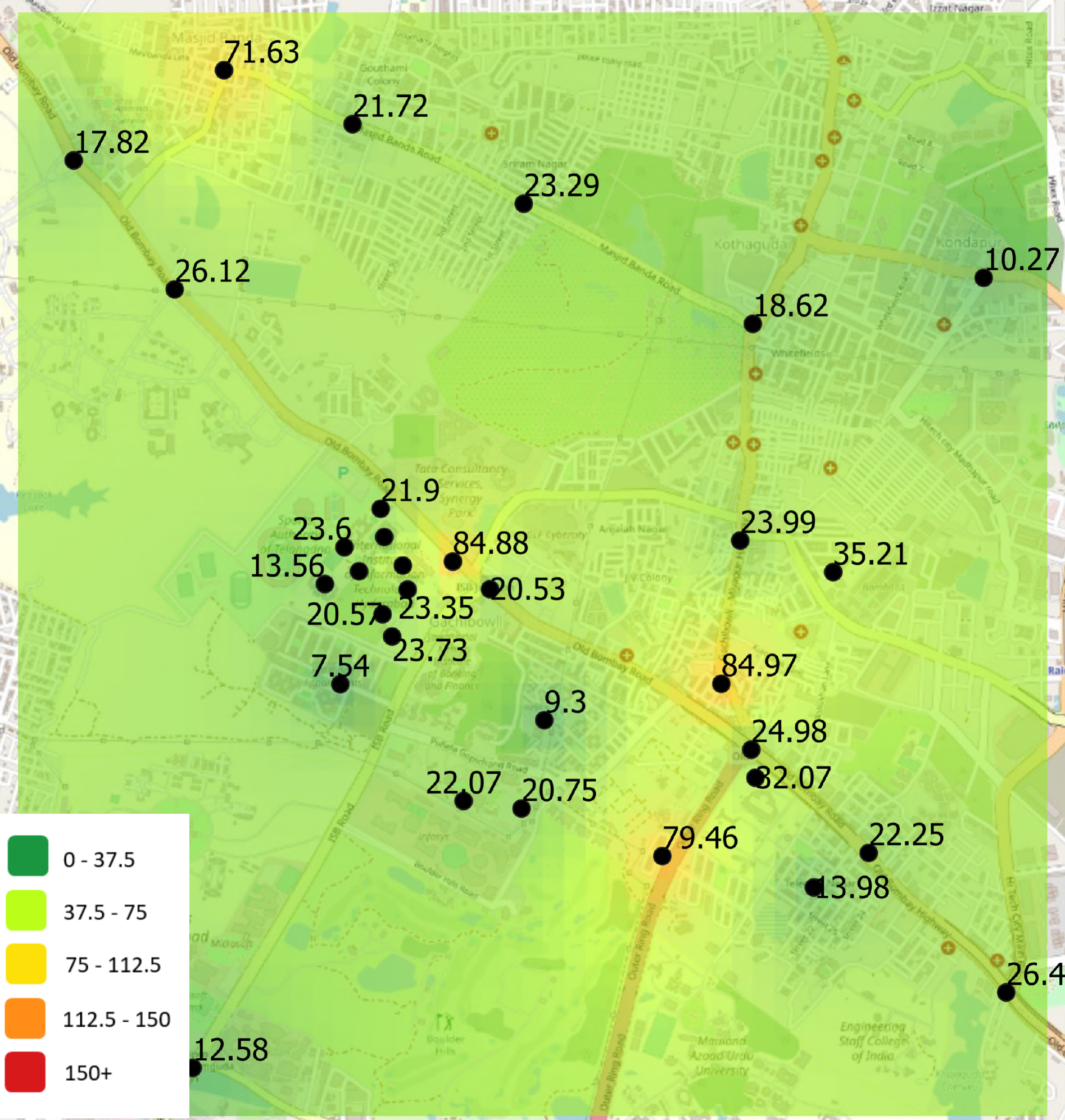}\label{IDW_Aug_a}}
\hfil
\subfloat[At 1400 hrs]{\includegraphics[width=0.3\linewidth]{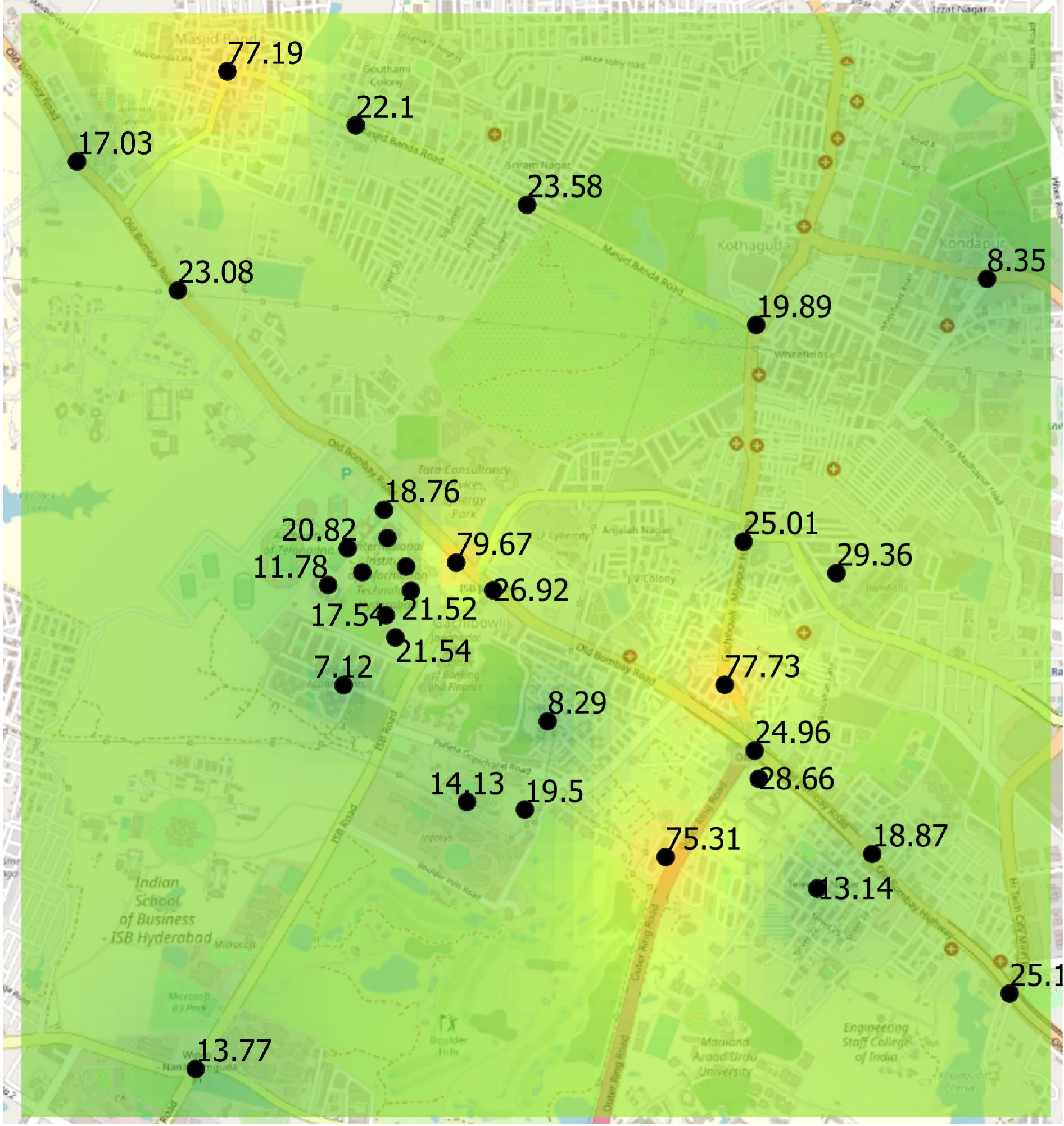}\label{IDW_Aug_b}}
\hfil
\subfloat[At 2100 hrs]{\includegraphics[width=0.3\linewidth]{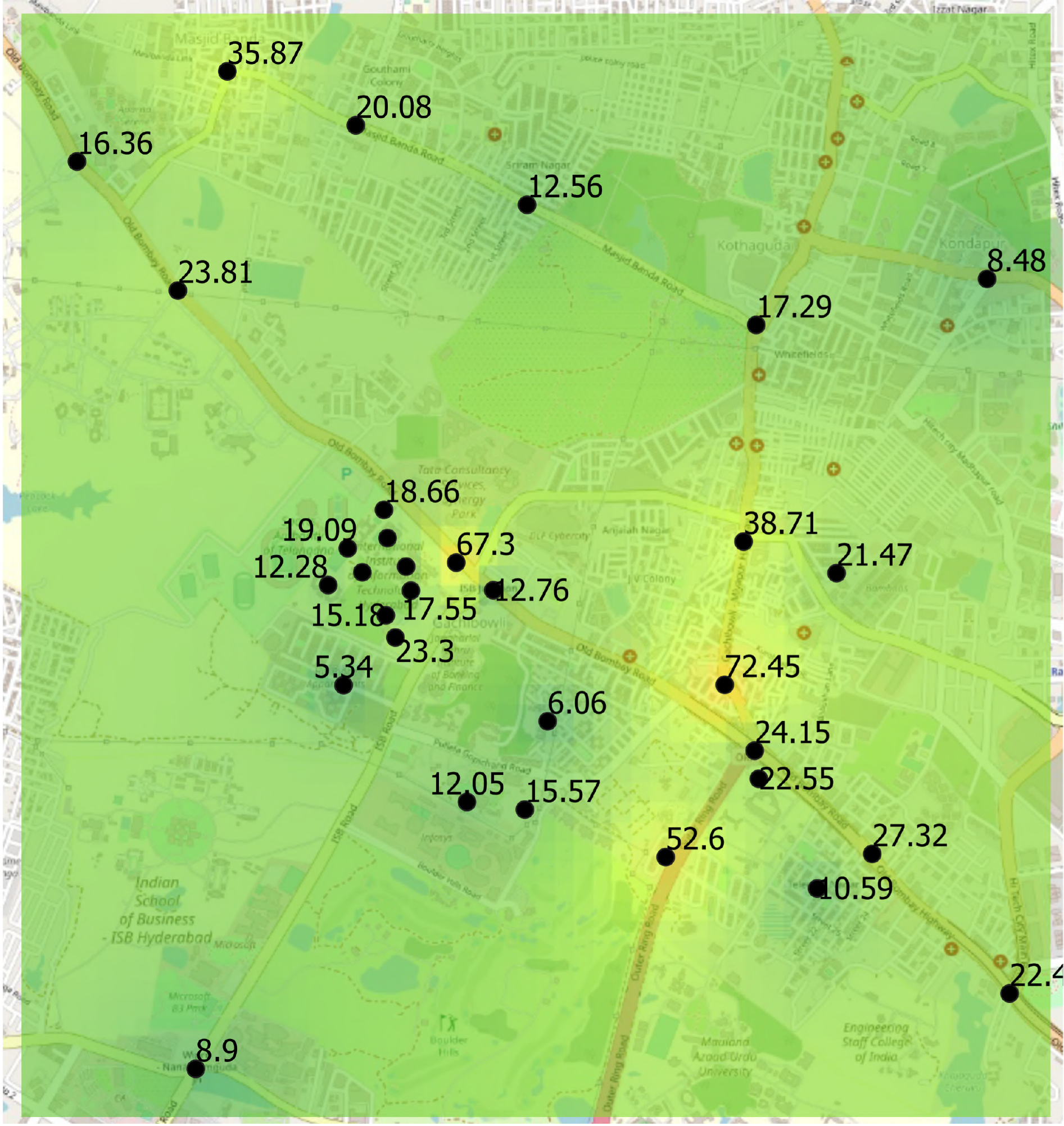}\label{IDW_Aug_c}}
\caption{Spatial interpolation of PM10 values in Monsoon (Aug. 2021) using IDW.}
\label{IDW_Aug}
\end{figure*}
%
\begin{figure*}[tbh]
\centering
\subfloat[At 1100 hrs]{\includegraphics[width=0.3\linewidth]{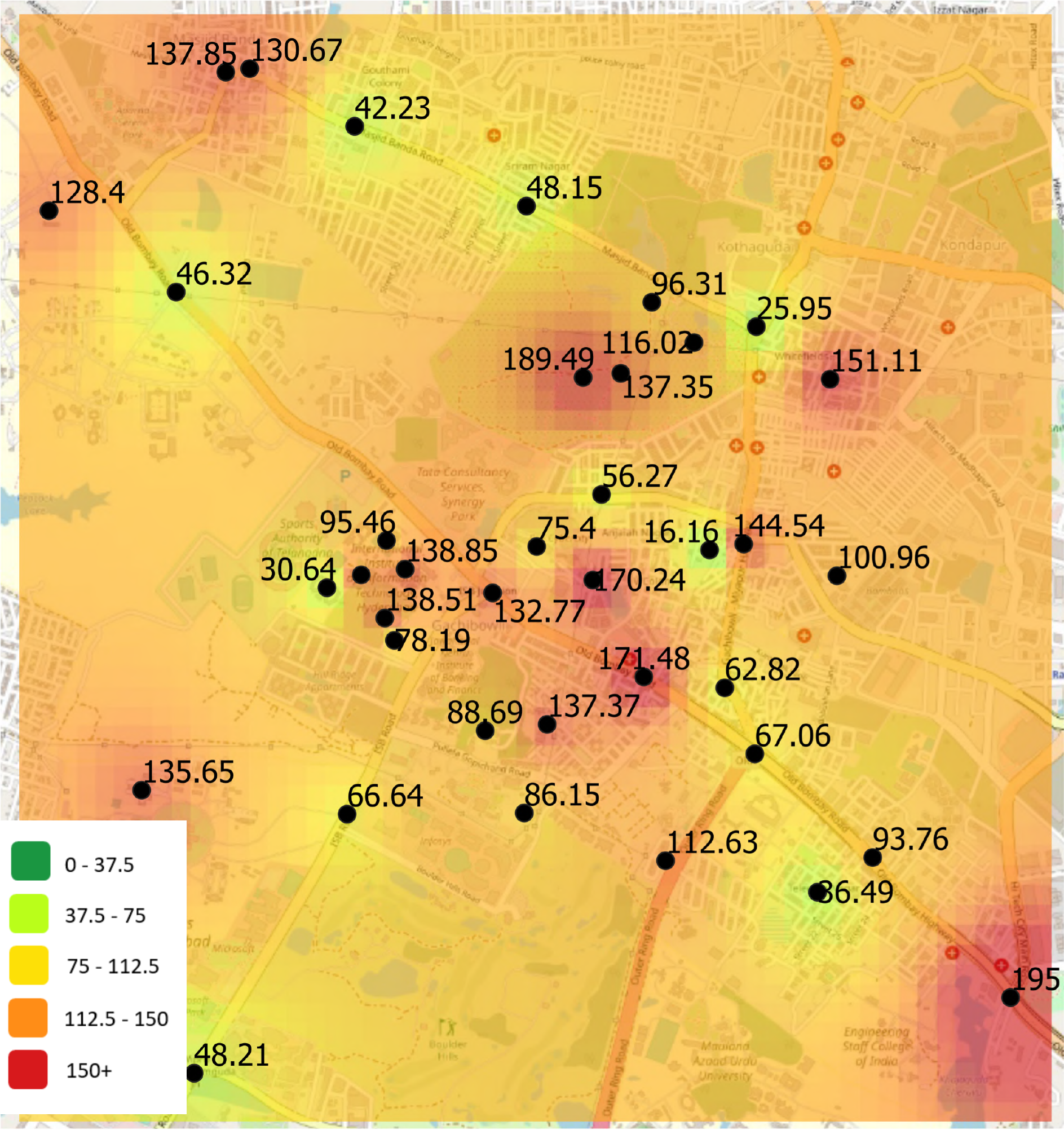}\label{IDW_DEC_a}}
\hfil
\subfloat[At 1400 hrs]{\includegraphics[width=0.3\linewidth]{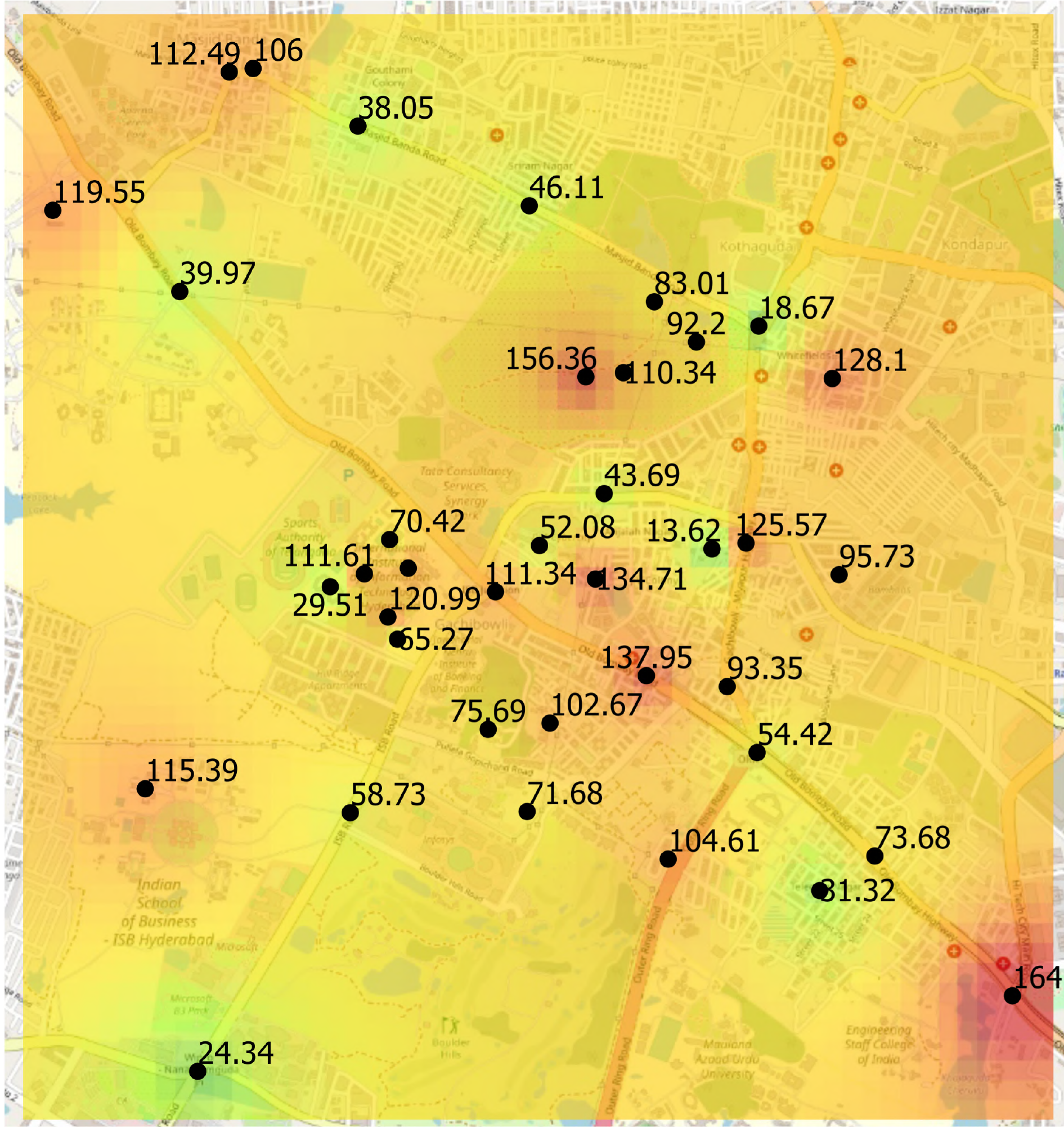}\label{IDW_DEC_b}}
\hfil
\subfloat[At 2100 hrs]{\includegraphics[width=0.3\linewidth]{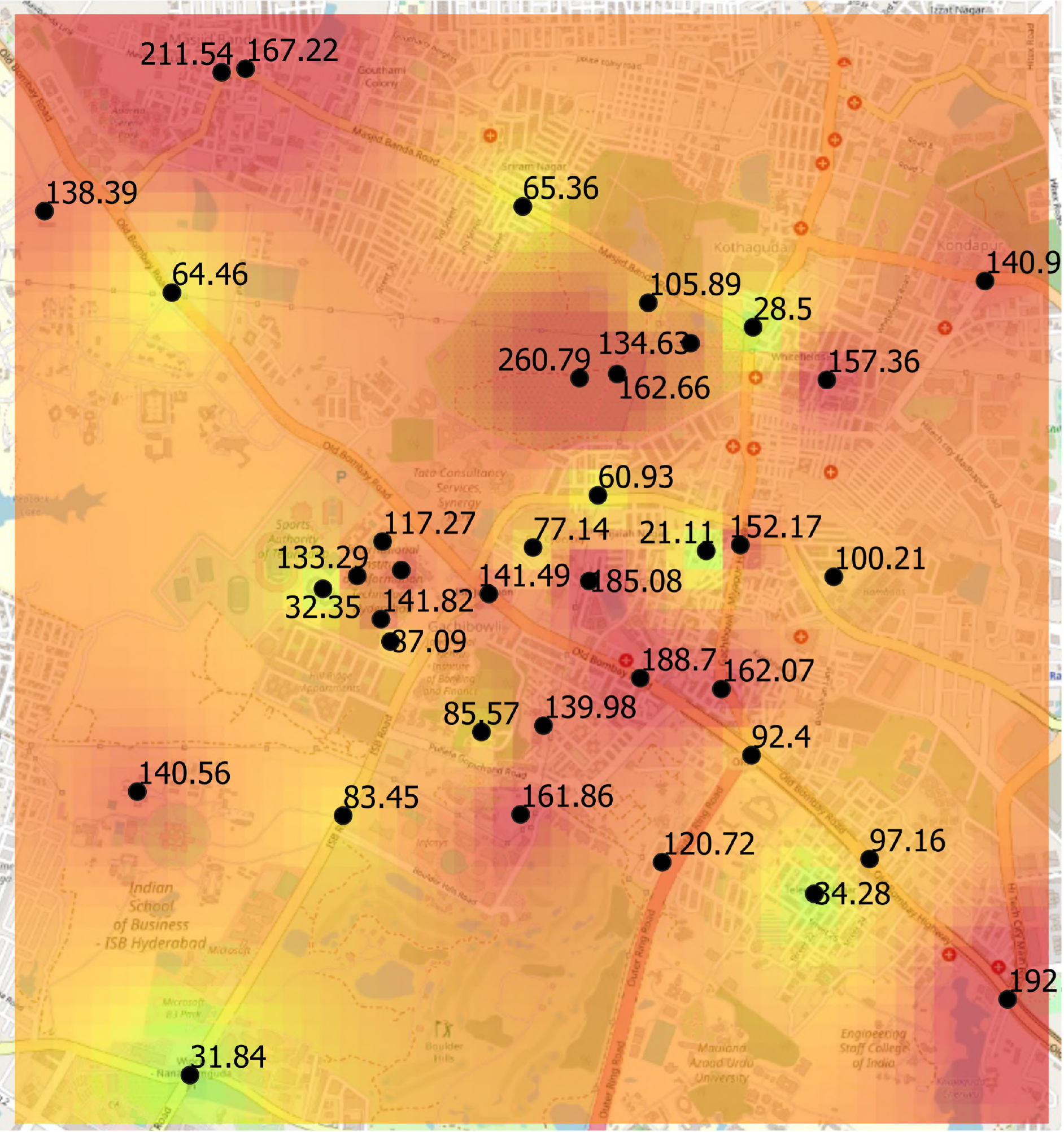}\label{IDW_DEC_c}}
\caption{Spatial interpolation of PM10 values in Winter (Dec. 2021) using IDW.}
\label{IDW_DEC}
\end{figure*}
%
\begin{figure*}[th!]
\centering
\subfloat[At 1100 hrs]{\includegraphics[width=0.3\linewidth]{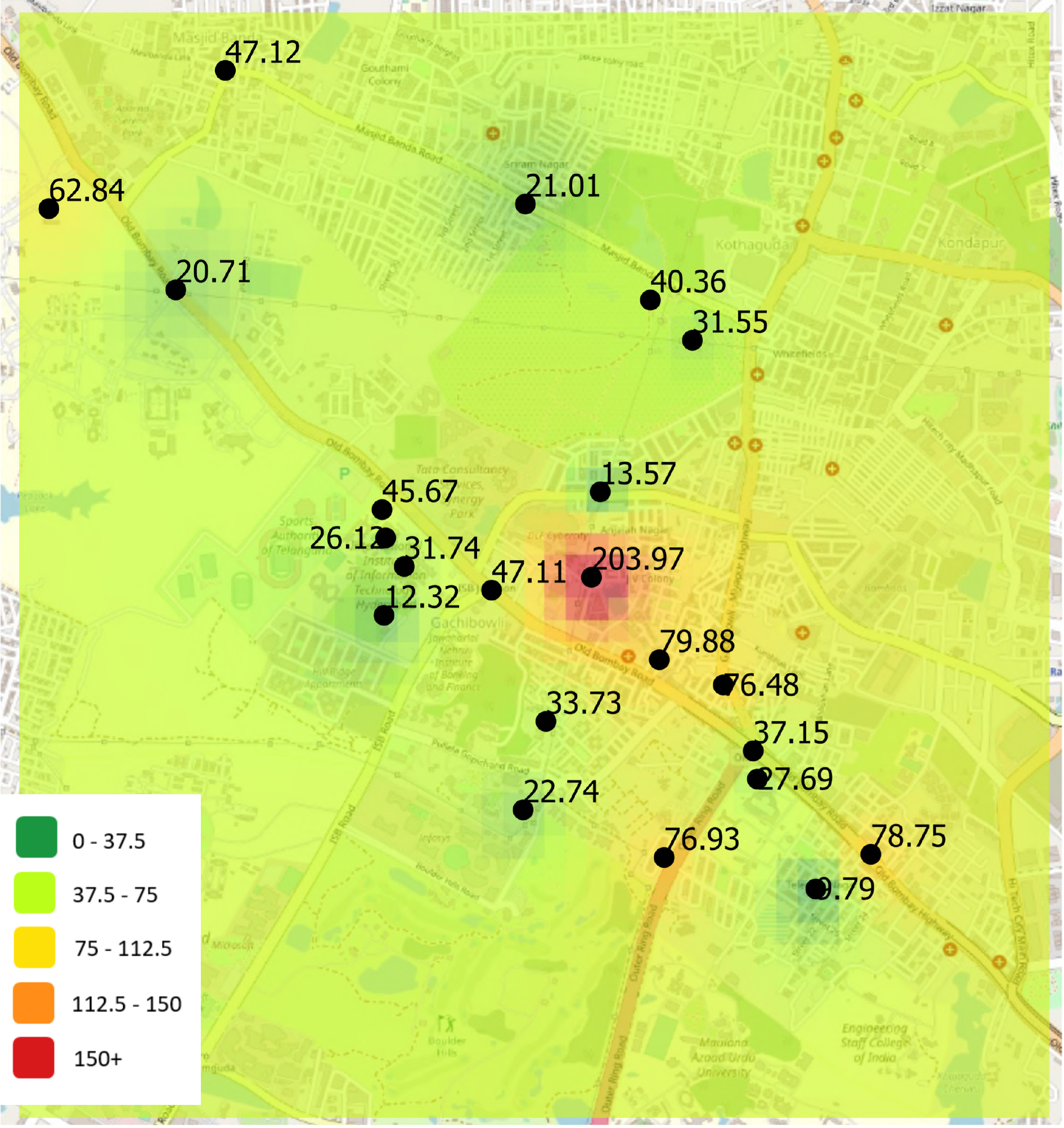}\label{IDW_May_a}}
\hfil
\subfloat[At 1400 hrs]{\includegraphics[width=0.3\linewidth]{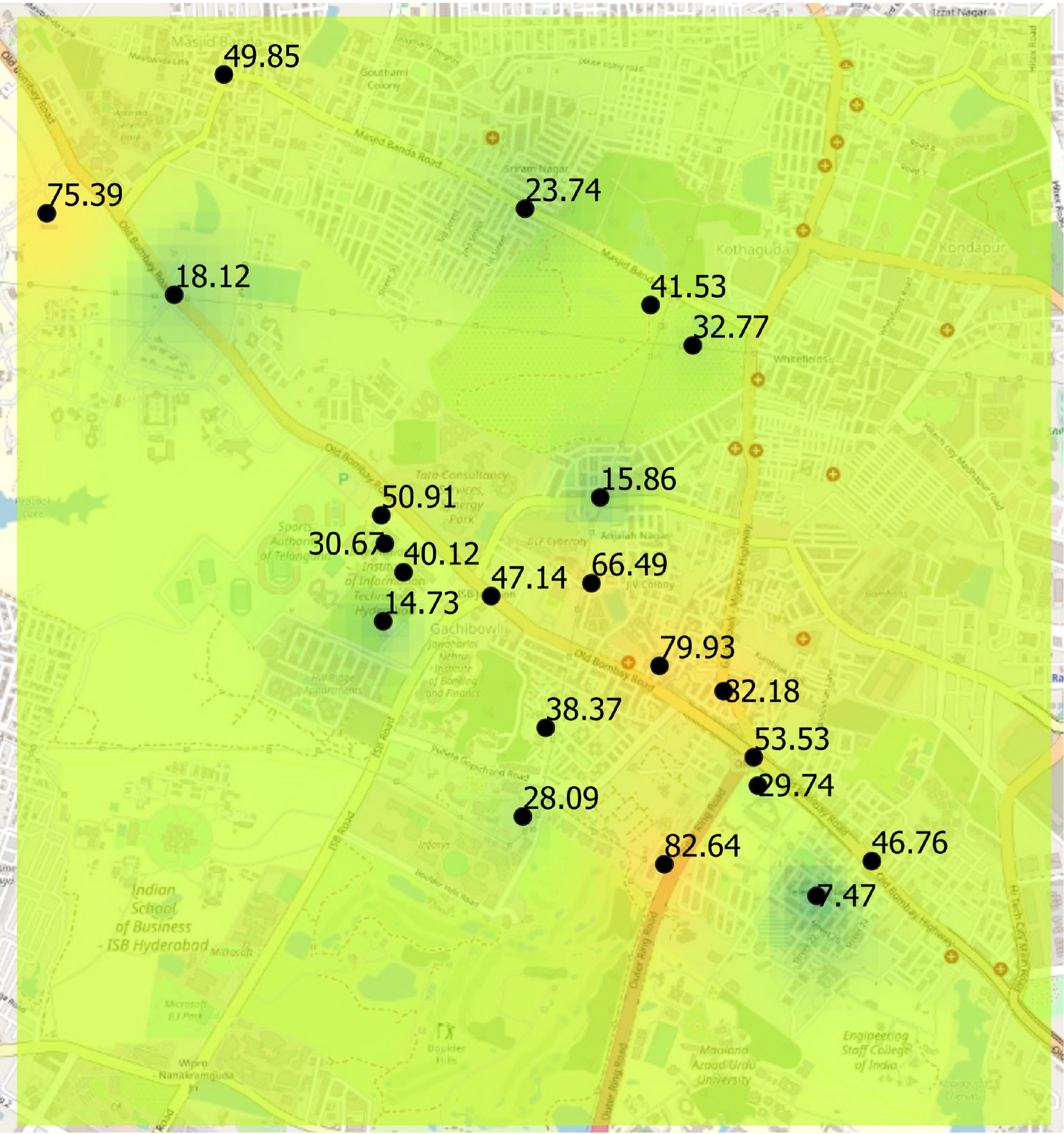}\label{IDW_May_b}}
\hfil
\subfloat[At 2100 hrs]{\includegraphics[width=0.3\linewidth]{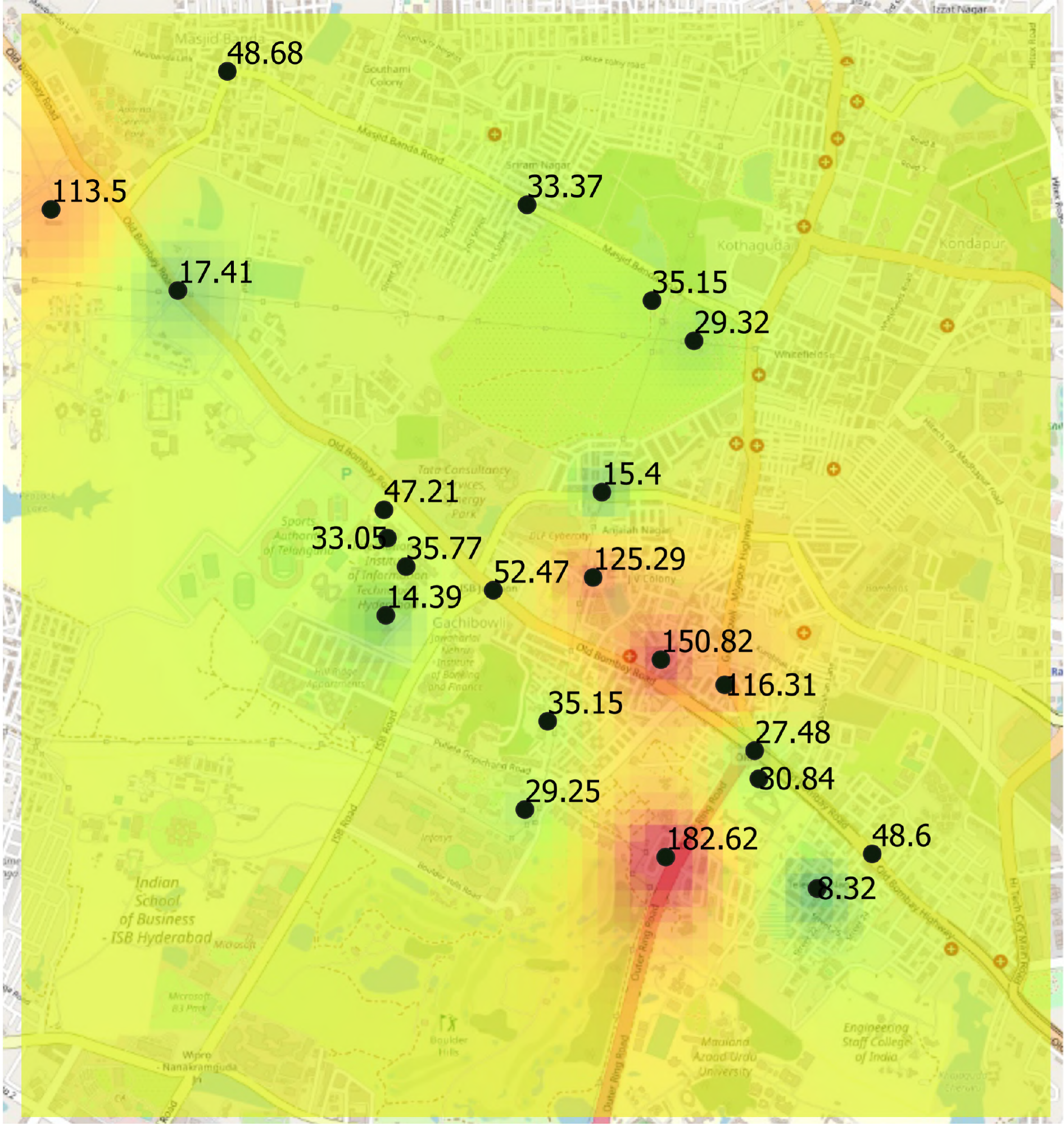}\label{IDW_May_c}}
\caption{Spatial interpolation of PM10 values in Summer (May 2022) using IDW.}
\label{IDW_May}
\end{figure*}
\subsection{Mean and Variance}
Fig. \ref{mean_var_pm10} shows the mean and variance of PM10 in monsoon (Aug. 2021), winter (Dec. 2021) and summer (May 2022). It can be observed that the mean and variance values are highest in winter and lowest in monsoon. This is expected as the surface temperature inversion (cold air near the ground and warm air on top) in winter trap PM near the ground. On the other hand, frequent rains during monsoons settle the PM, reducing their concentration in the air. It can also be observed that there is a lot of variation in the mean and the variance of the PM values among the various devices in the same geographical region, demonstrating the need for dense deployment to understand street-level pollution.

In Fig. \ref{mean_var_pm10}, the three devices with the highest mean PM10 values among the 49 devices are AQ23 (Traffic junction), AQ20 (Green region), and AQ16 (Roadside), while the three devices with the lowest mean PM10 values are AQ11 (Residential area), AQ43 (Roadside), and AQ22 (Roadside). Devices like AQ23 near the traffic junctions have high PM10 exposure due to heavy traffic. Similarly, devices like AQ16 near traffic lights have sluggish traffic flow leading to high mean PM10 concentrations. AQ20 is placed in high vegetation area but still shows a high mean due to ongoing construction activities in the region. Among the ones with low mean values, AQ11 is placed in a residential area with fewer anthropogenic activities. Similarly, AQ43 and AQ22 are otherwise placed on the roadside but still experience low mean PM10 due to the free flow of traffic and less anthropogenic activities.
\subsection{Spatial Interpolation}
\vspace{-0.1cm}
Inverse distance weighting (IDW) is used for spatial interpolation, one of the most popular spatial interpolation techniques. IDW follows the principle that closer devices will have more impact than farther devices \cite{IDW_Book}. A linearly weighted combination of the measured values at the devices is used to estimate the parameters at the nearest location. The weights are a function of the inverse distance between the device's location and the estimate's location. 

Figs. \ref{IDW_Aug}, \ref{IDW_DEC} and  \ref{IDW_May} are the IDW-based interpolation maps for PM10 in monsoon (Aug 21), winter (Dec 21) and summer (May 22), respectively. For all three seasons, the interpolation results are shown at three different times of the day, 1100 hrs, 1400 hrs and 2100 hrs, based on hourly averaged PM values. Similar to the observations from Fig. \ref{mean_var_pm10}, it can also be observed in these figures that the PM concentrations are lowest in monsoon and highest in winter. 

%
\begin{figure*}[t!]
\centering
\setkeys{Gin}{width=\linewidth}
\begin{tabularx}{\linewidth}{XXXX}
    \centering
    \includegraphics[width=0.94\linewidth]{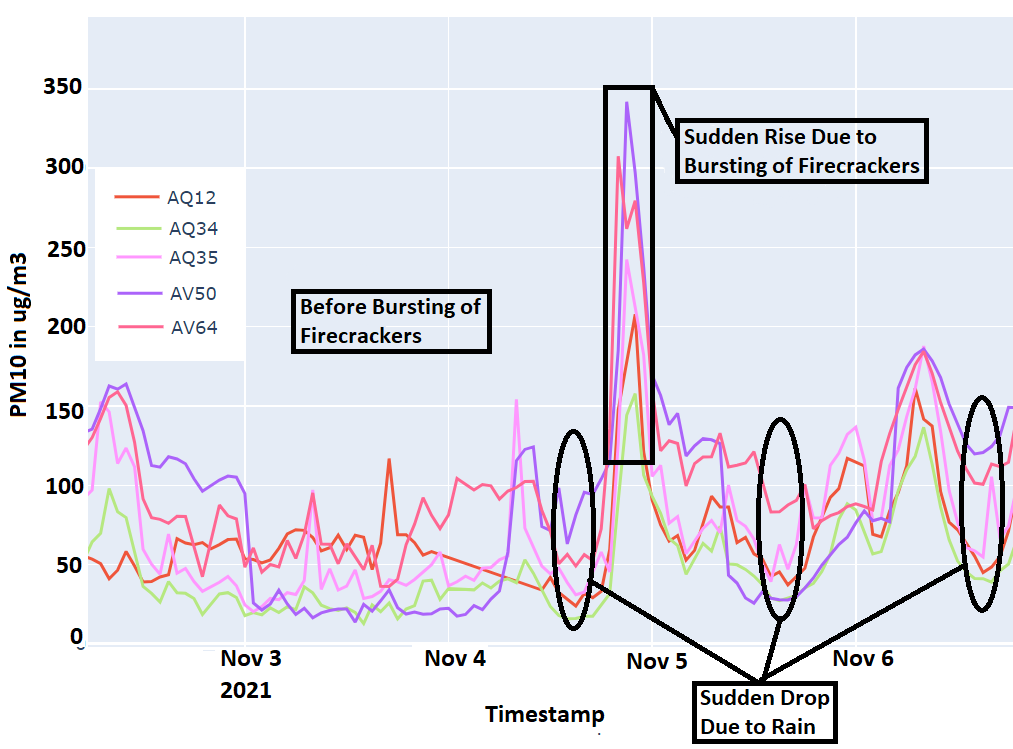}
    \caption{Time Series of PM10 (1-Hourly Average) showing the rise in PM10 due to bursting of firecrackers during Diwali.}
    \label{fig:TimeSeries_Diwali}
&
    \centering
    \includegraphics[width=0.72\linewidth]{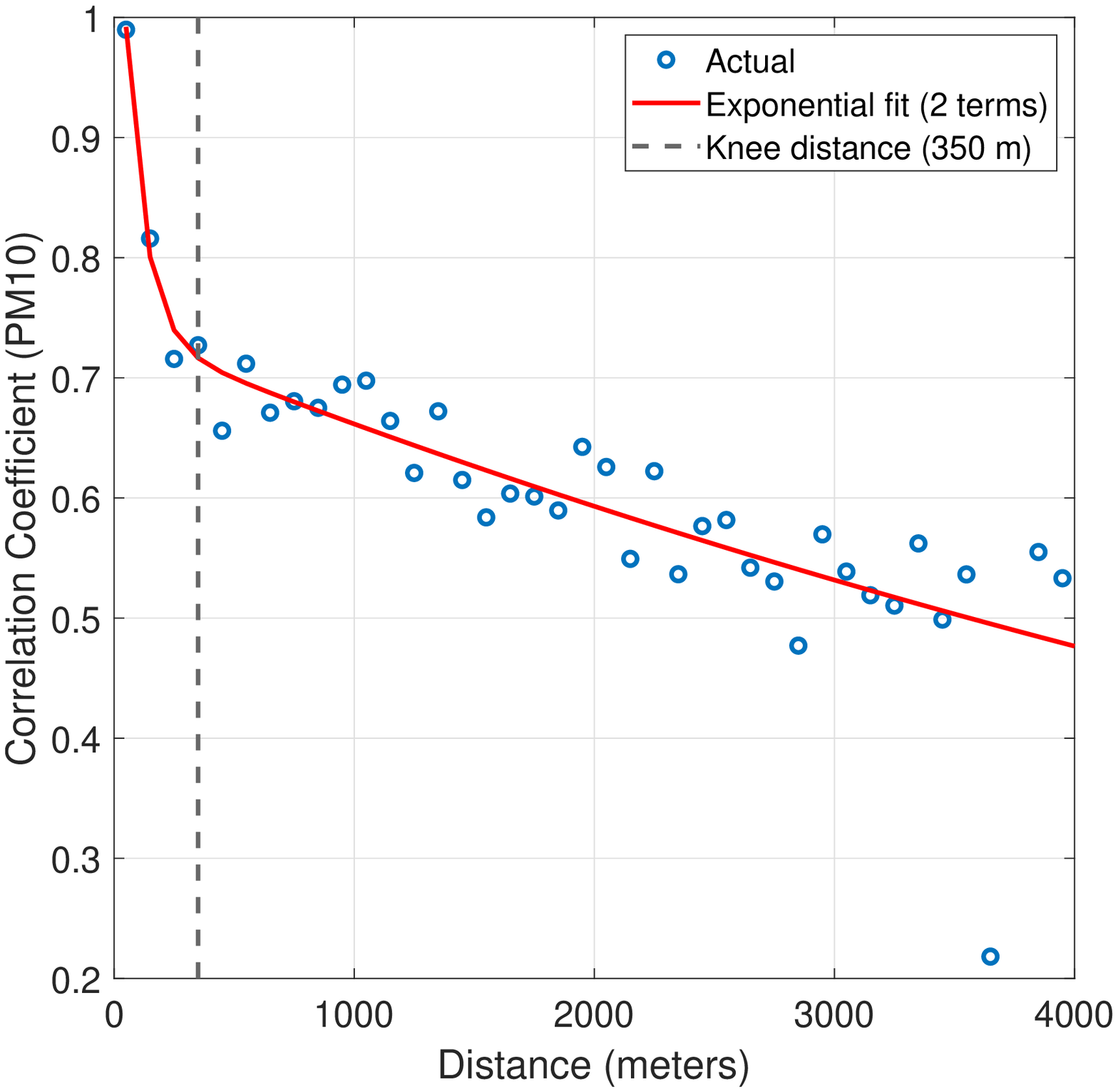}
    \caption{Correlation coefficient of PM10 \textit{w.r.t} distance between the deployed devices to find the optimum distance between two deployment locations.}
    \label{fig:correlation}
\end{tabularx}
\vspace{-0.6cm}
\end{figure*}

It can be observed from Figs. \ref{IDW_Aug}, \ref{IDW_DEC} and  \ref{IDW_May} that PM concentration was high at 1100 hrs and 2100 hrs and low at 1400 hrs. At 1100 hrs, PM concentration was high, primarily due to heavy traffic. As the day progresses, the density of traffic decreases and the PM concentration decrease at 1400 hrs. However, with the onset of night, PM concentrations can be seen as increased at 2100 hrs, falling in peak traffic hours.



%
\subsection{Event Driven Variation Analysis}
Diwali, also known as the festival of lights, is celebrated during the start of the winter. As part of this five-day festival, people burst large numbers of firecrackers in the late evening of the third day of Diwali (4 Nov. 2021). The bursting of firecrackers leads to a significant increase in PM values during those times. Fig. \ref{fig:TimeSeries_Diwali} shows a time series plot of hourly averaged PM10 values for a few devices over a few days around Diwali. A few critical observations can be made from this figure. First, there was a sudden drop in the PM10 values on the afternoon of 4\textsuperscript{th} Nov. because of rain. The same has been observed on 5\textsuperscript{th} and 6\textsuperscript{th} Nov. afternoons. 
Second, a clear peak is observed for all the devices during the late evening on 4 Nov. 2021, roughly after 2000 hrs. For example, the PM10 values in AV64 increased from 40 to 307 before and after bursting crackers. This peak can be attributed to the widespread bursting of firecrackers during the festive celebrations. 
Third, it can be observed that the PM10 concentrations decrease sharply after a few hours, indicating that the rise was temporary and activity driven.\looseness=-1 

\begin{figure*}[tbh]
\centering
\subfloat[At 1700 hrs]{\includegraphics[width=0.32\linewidth]{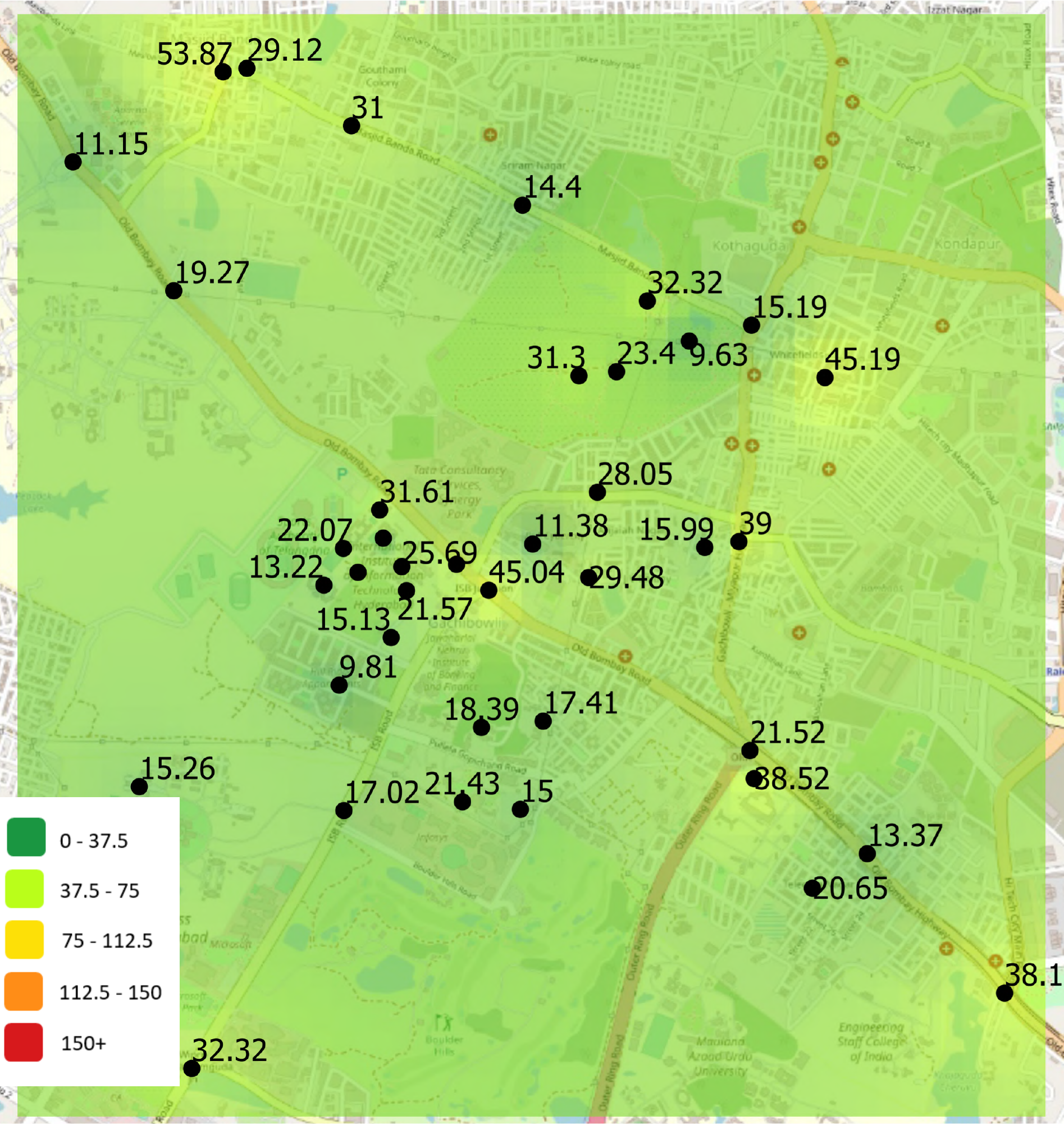}\label{IDW_Diwali_a}}
\hfill
\subfloat[At 2100 hrs]{\includegraphics[width=0.32\linewidth]{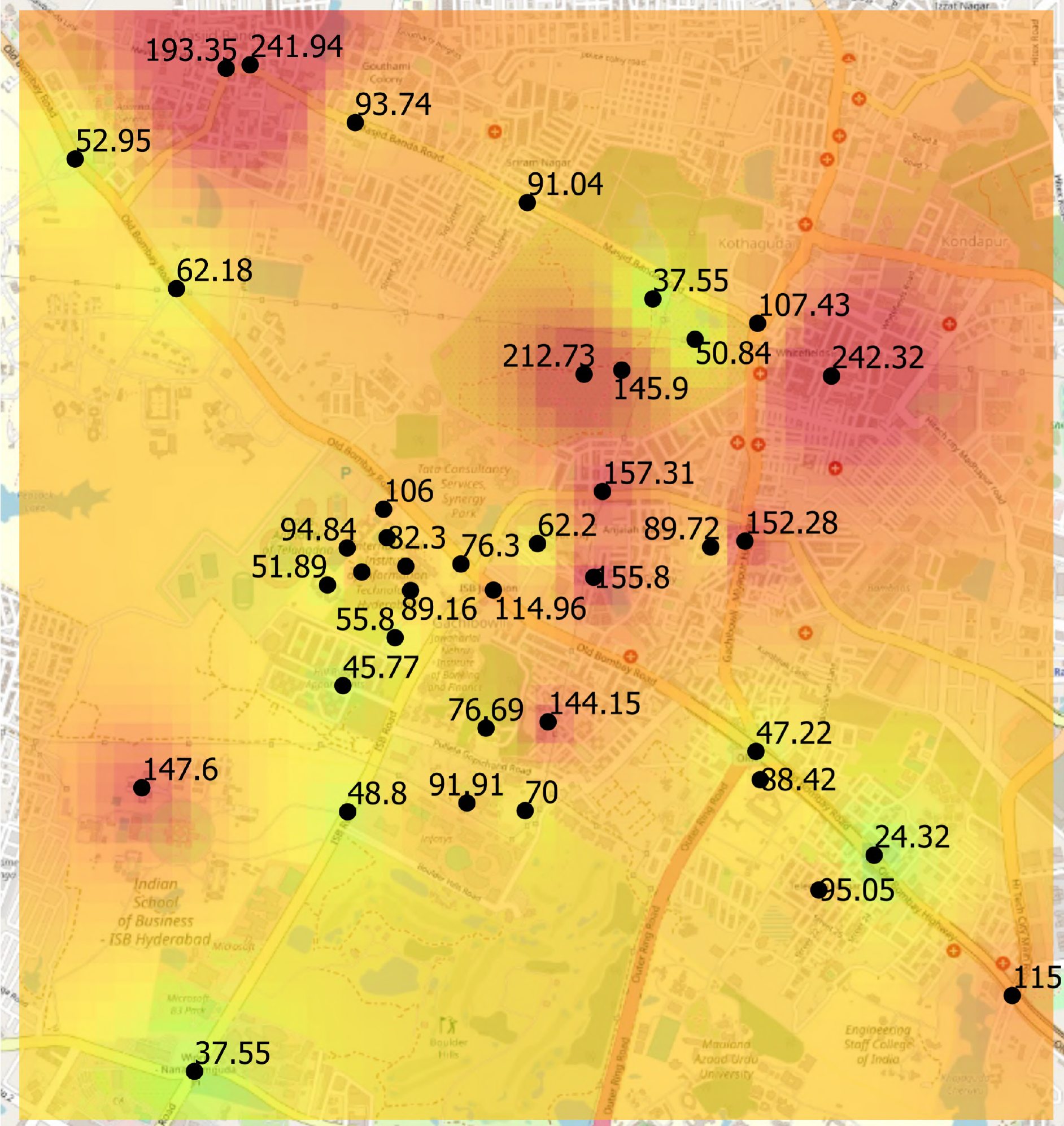}\label{IDW_Diwali_b}}
\hfill
\subfloat[At 2300 hrs]{\includegraphics[width=0.334\linewidth]{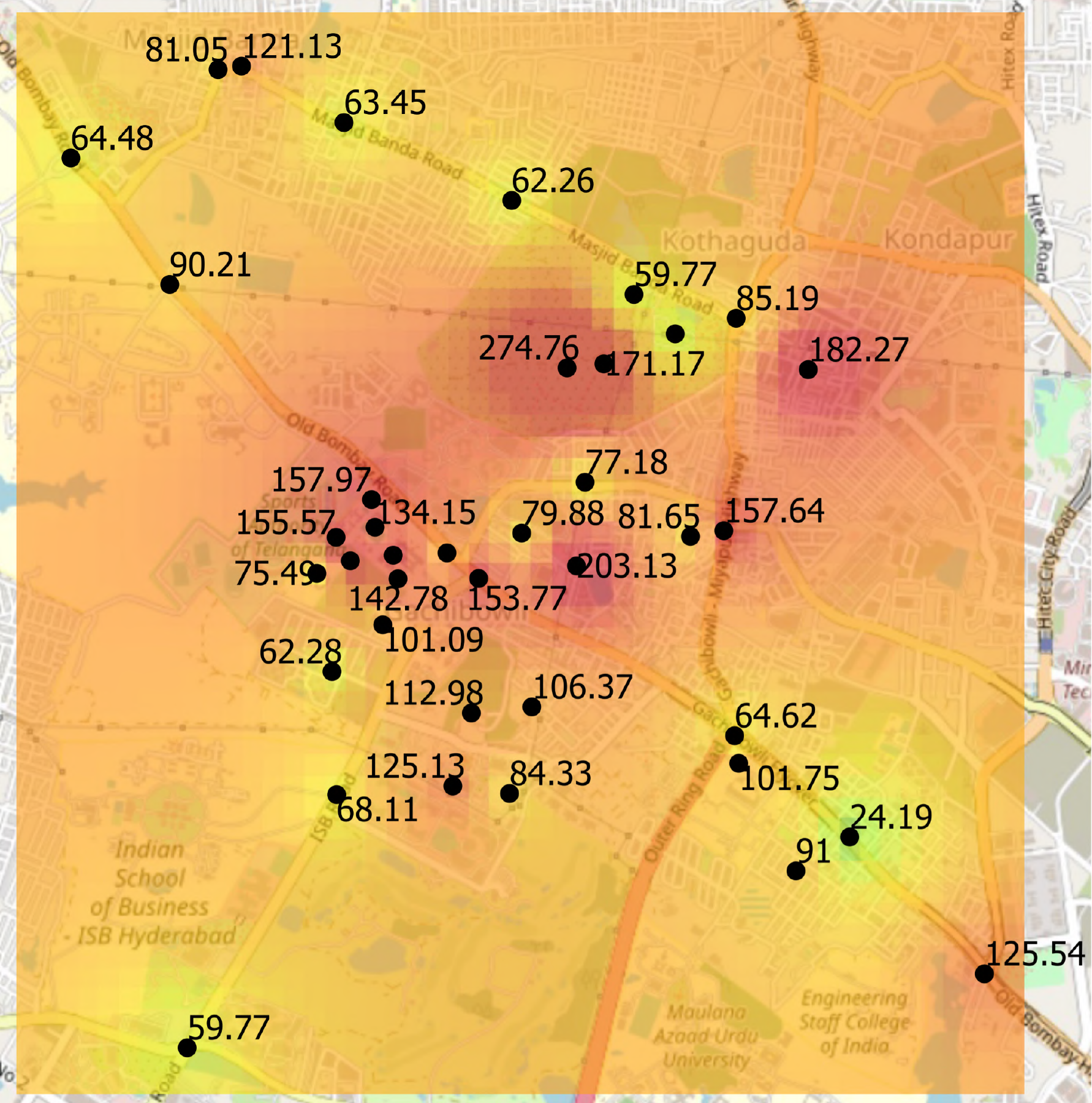}\label{IDW_Diwali_c}}
\caption{Spatial interpolation of PM10 values from densely deployed devices during Diwali 2021 using IDW.}
\label{IDW_Diwali}
\end{figure*}
\begin{figure*}[tbh]
\centering
\subfloat[At 1700 hrs]{\includegraphics[width=0.32\linewidth]{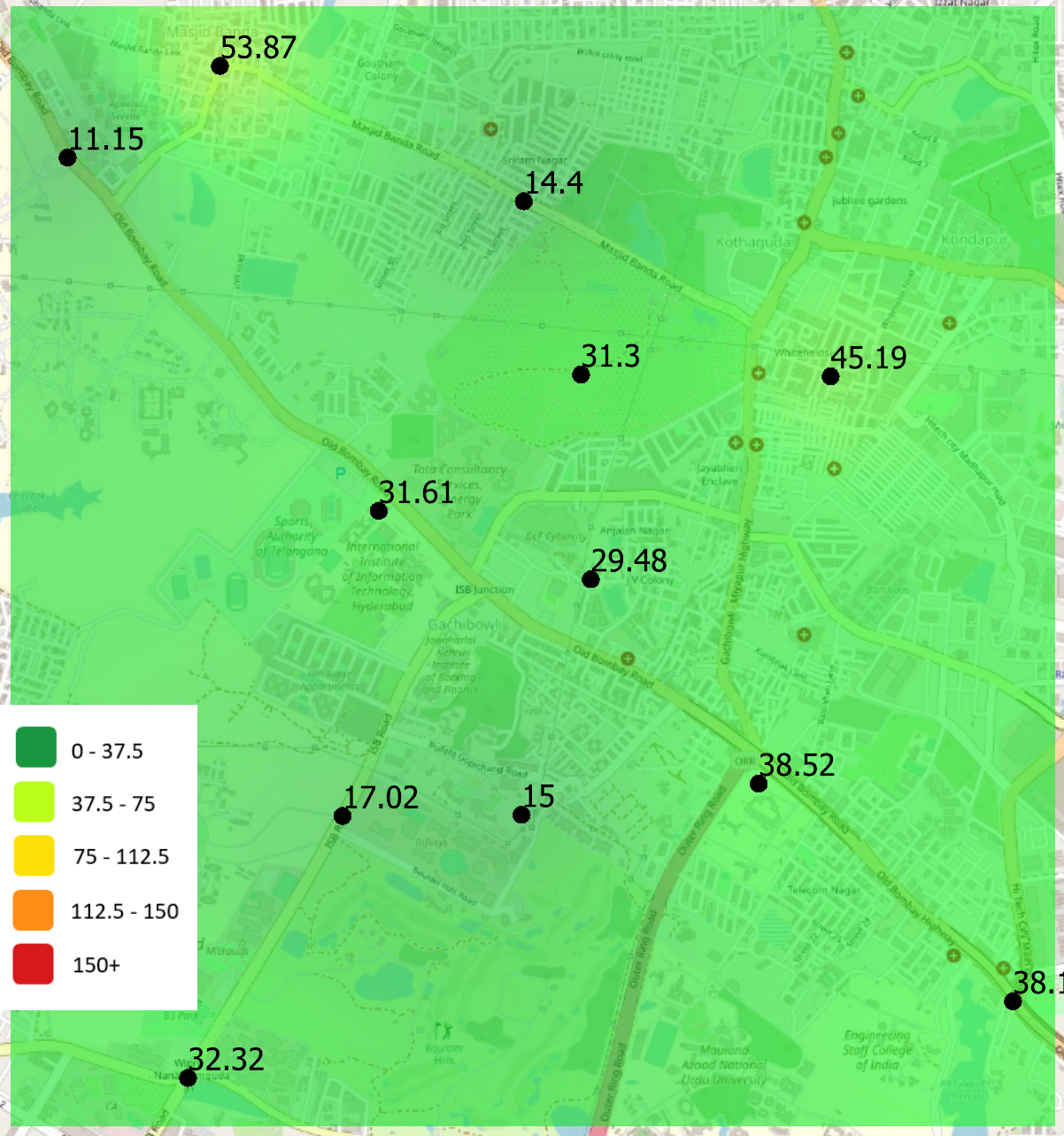}\label{IDW_Diwali_sparse_a}}
\hfill
\subfloat[At 2100 hrs]{\includegraphics[width=0.32\linewidth]{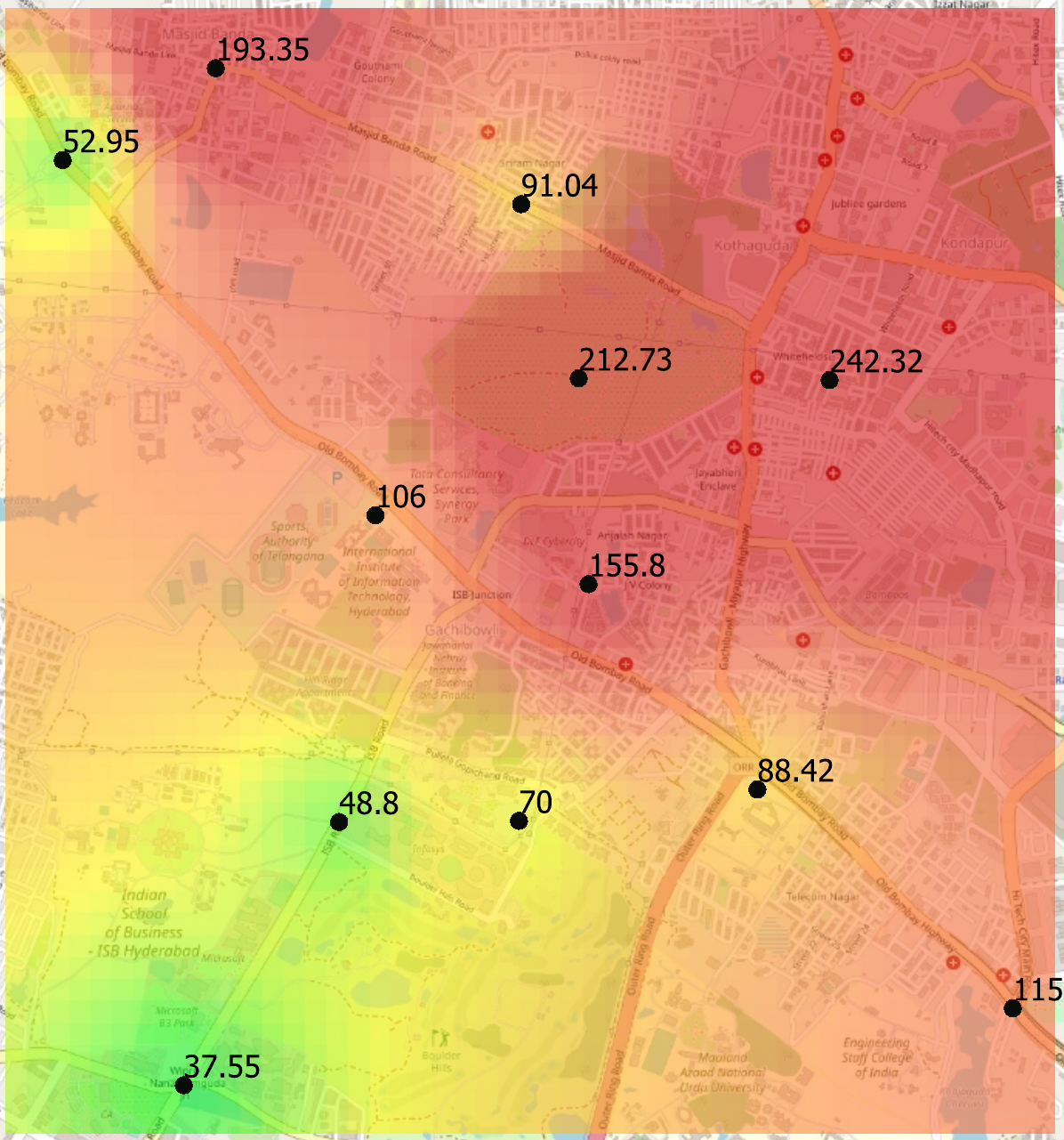}\label{IDW_Diwali_sparse_b}}
\hfill
\subfloat[At 2300 hrs]{\includegraphics[width=0.32\linewidth]{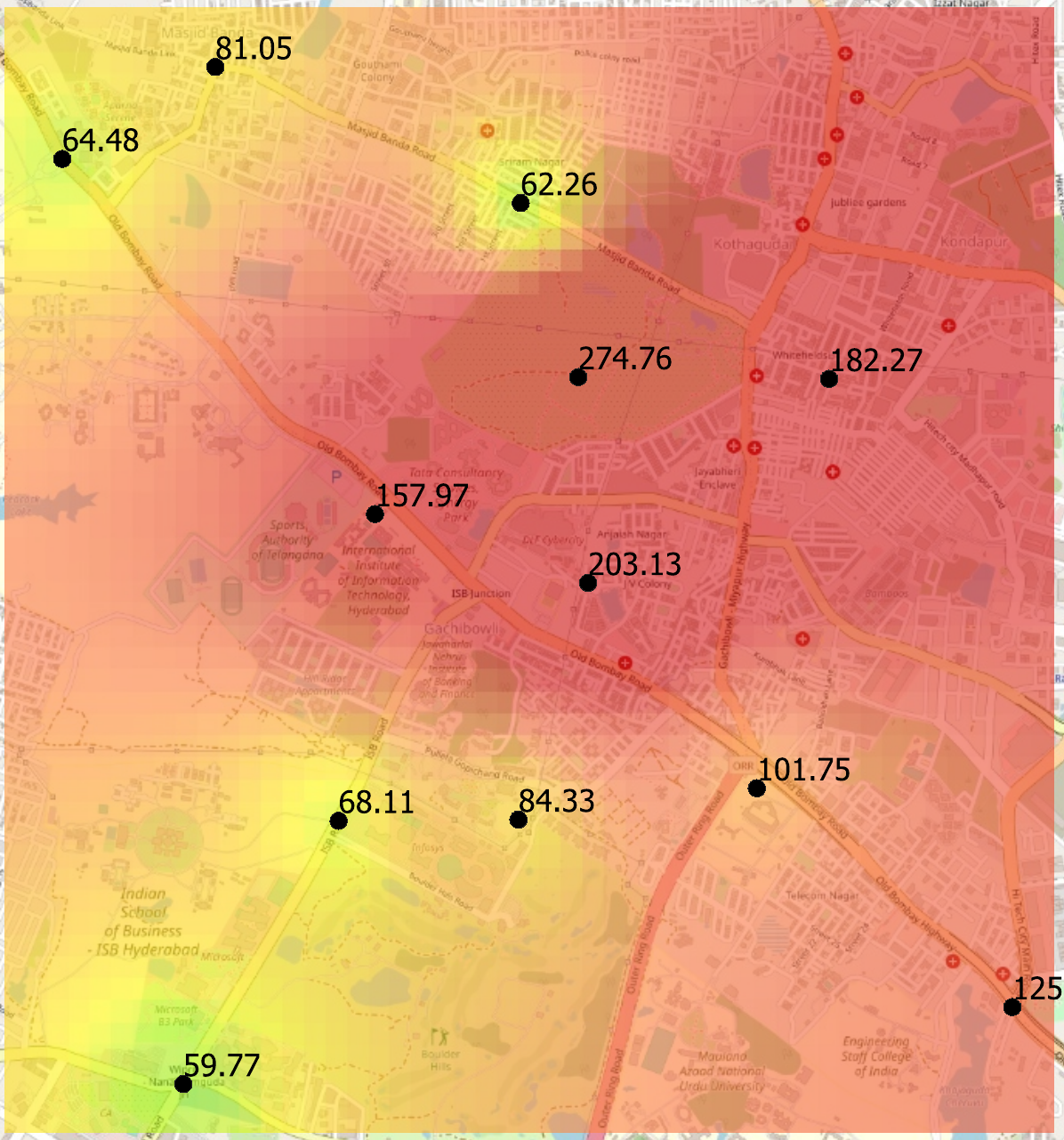}\label{IDW_Diwali_sparse_c}}
\caption{Spatial interpolation of PM10 values from 12 sparsely deployed devices during Diwali 2021 using IDW.}
\label{IDW_Diwali_sparse_12}
\end{figure*}
\begin{figure*}[th!]
\centering
\subfloat[At 1700 hrs]{\includegraphics[width=0.32\linewidth]{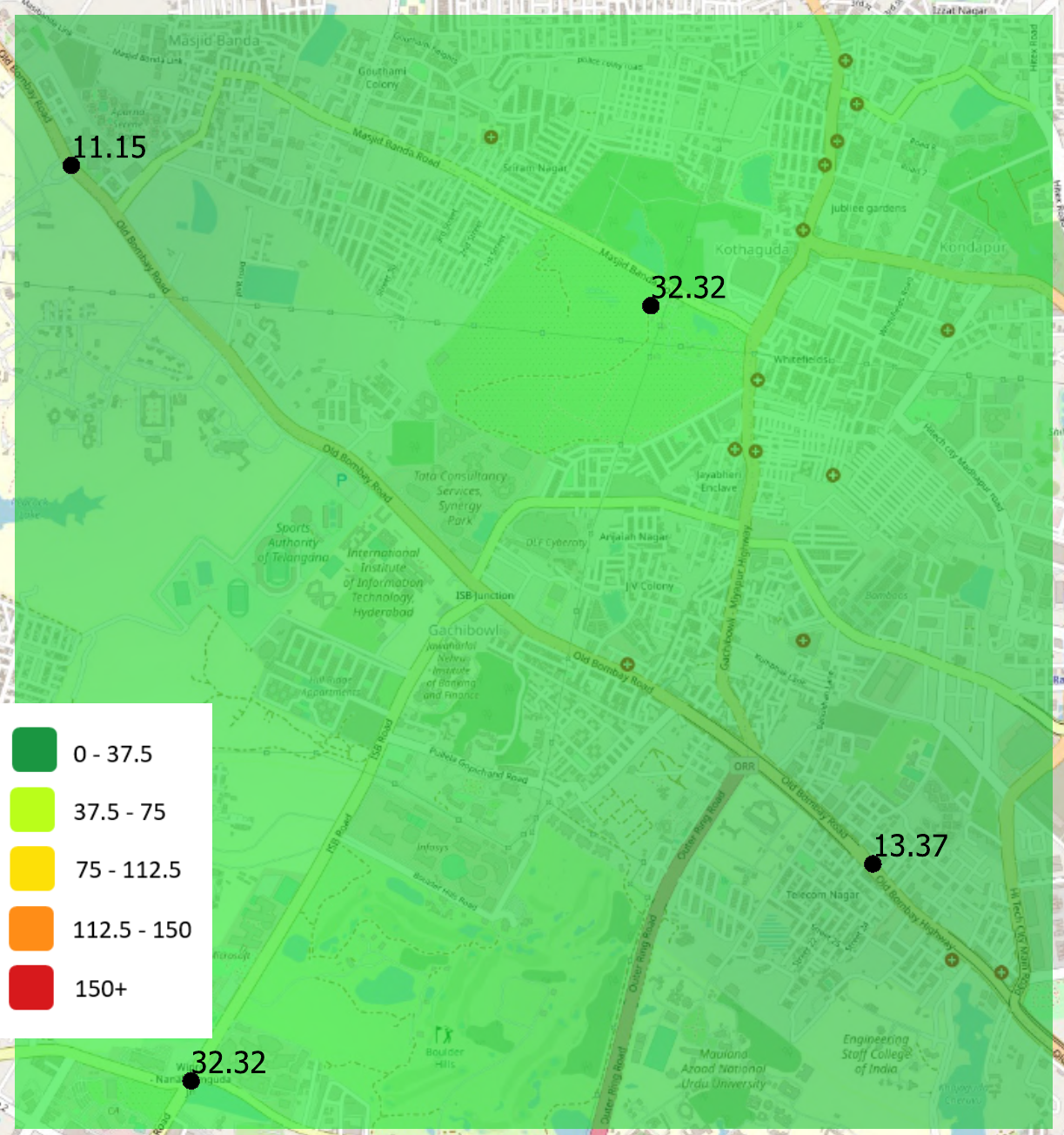}\label{IDW_Diwali_sparse4_a}}
\hfill
\subfloat[At 2100 hrs]{\includegraphics[width=0.32\linewidth]{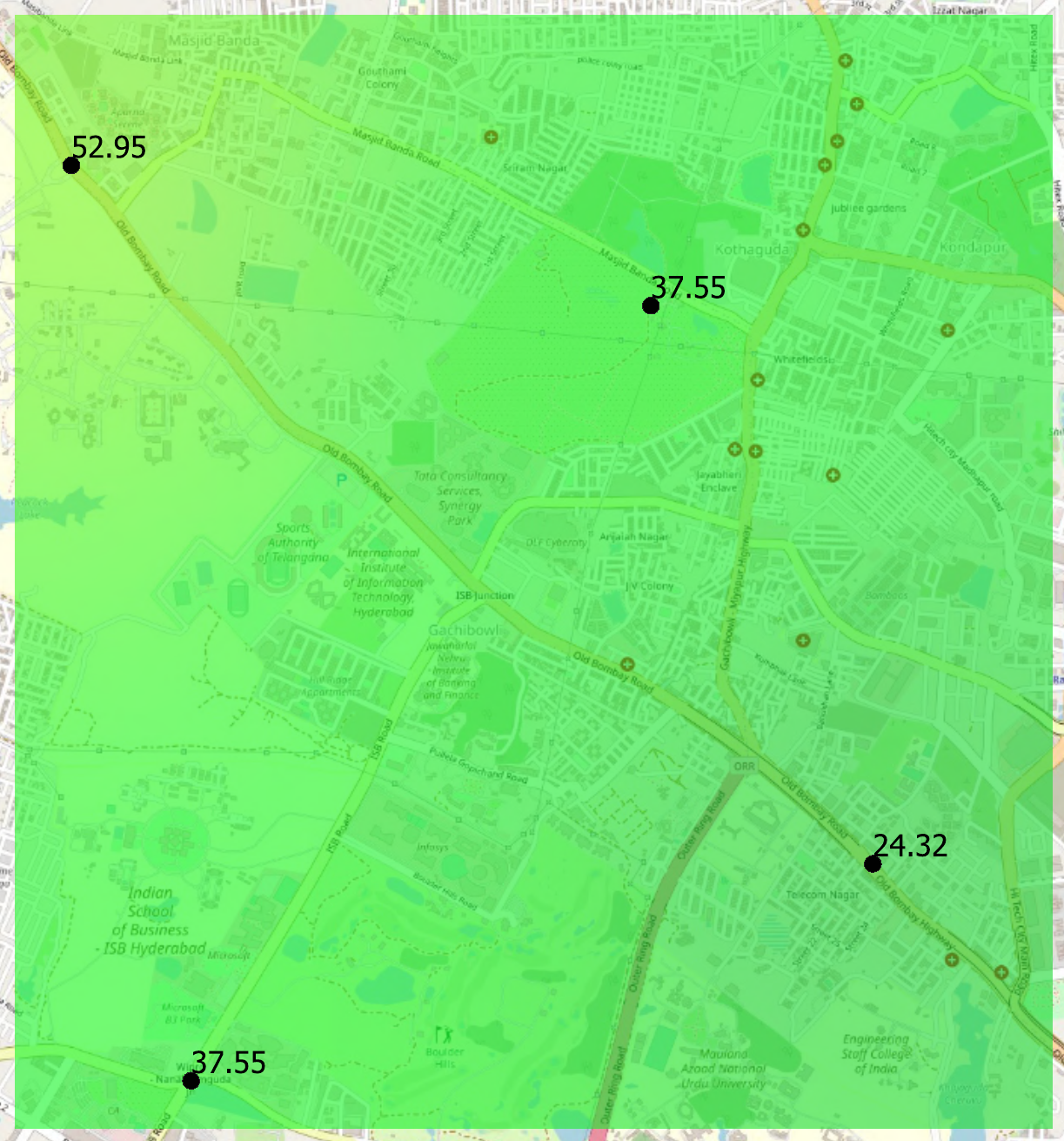}\label{IDW_Diwali_sparse4_b}}
\hfill
\subfloat[At 2300 hrs]{\includegraphics[width=0.32\linewidth]{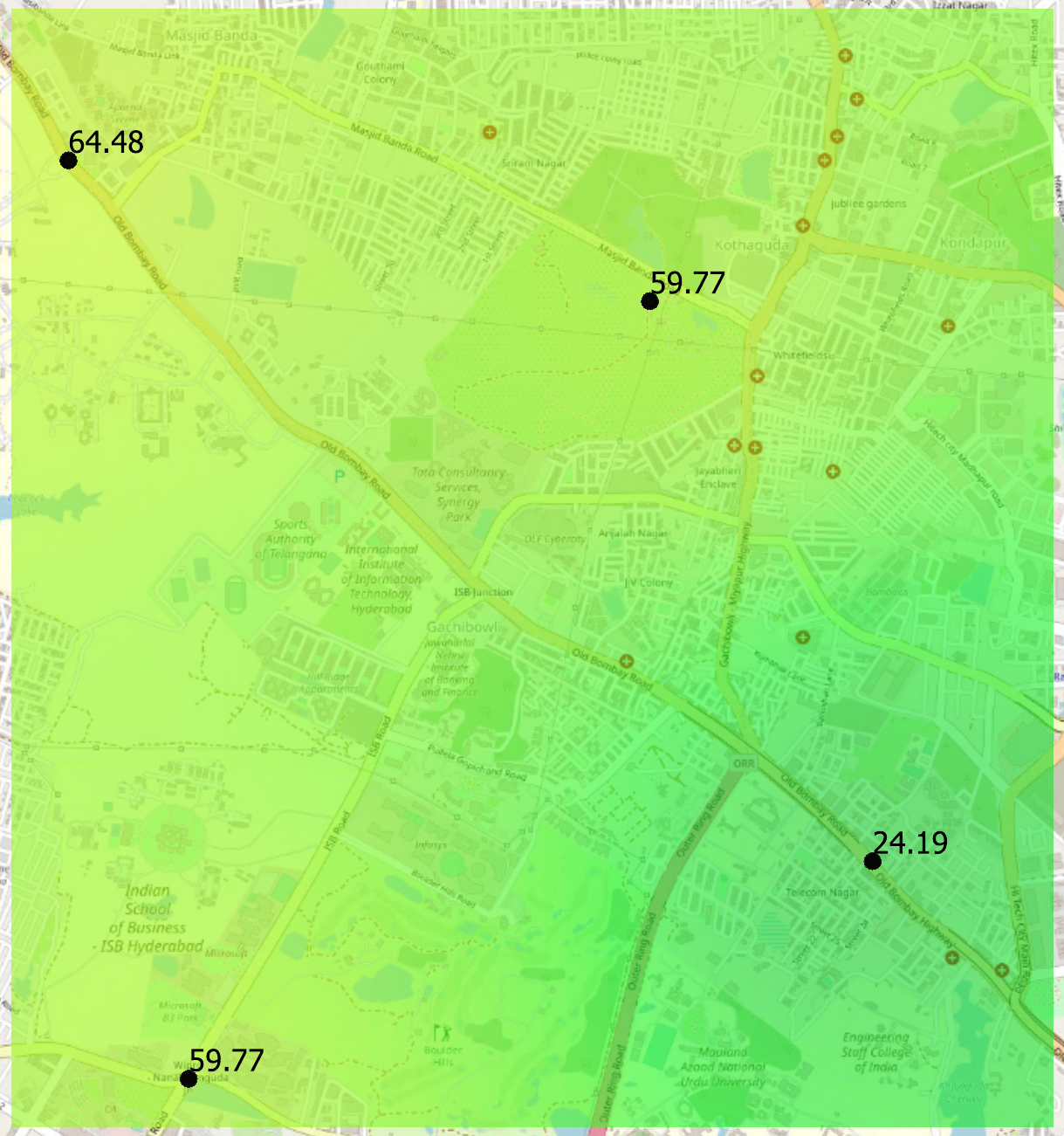}\label{IDW_Diwali_sparse4_c}}
\caption{Spatial interpolation of PM10 values from 4 sparsely deployed devices during Diwali 2021 using IDW.}
\label{IDW_Diwali_sparse}
\end{figure*}

Further, we see the effect of sparse deployment on the event-driven analysis. Figs.\ref{IDW_Diwali}, \ref{IDW_Diwali_sparse_12} and  \ref{IDW_Diwali_sparse} show the IDW-based interpolation maps for PM10  using all 49 devices and sparse deployment of 12 and 4 devices, respectively, at different time instances on 4 Nov. 2021. For sparse deployment, 4 and 12 devices were chosen randomly with the constraint that they should not form a cluster and should also cover different types of regions. It can be seen from Fig. \ref{IDW_Diwali} that the interpolation plot with all 49 devices can identify the event as well as the local hotspots of pollution. Although the interpolation in Fig. \ref{IDW_Diwali_sparse_12} (with 12 devices within $4\,\text{km}^2$) can identify the event but is not able to identify the hotspots. On the other hand, the interpolation (with 4 devices within $4\,\text{km}^2$) misses the event entirely and can be misleading. The root mean square error (RMSE) for different number of deployed devices (12 and 4 devices) is calculated. The RMSE for sparse deployments like 4 devices (59.29) is significantly more compared to 12 devices (32.47).\looseness=-1

%
\subsection{Correlation Analysis}
Correlation is a type of bivariate analysis that evaluates the direction and strength of an association between two variables \cite{Correlation}. Kendall's tau method is used as it does not require any presumptions on the data and suits the work in this study. The correlation coefficient's value ranges from -1 to +1 depending on the strength of the association. Kendall's correlation coefficients $\tau$ between the 49 sensor devices have been calculated using hourly averaged PM10 samples.

A two-term exponential fit is obtained on the correlation values when plotted against the distance between the devices. The fitted model can be written as
\begin{equation}
f(x) = a\,e^{b\,x} + c\,e^{d\,x},
\label{two-term exponential}
\end{equation}
where $a=0.4801, b=-0.0124, c=0.7380$ and $d=-0.0001$ are the coefficients of the best fit for PM10. Fig. \ref{fig:correlation} shows the correlation of PM10 plotted against distance. It can be observed that the change in the correlation coefficient under 350 meters is significantly large, after which the decline is gradual. The $\tau$ change rate between 0 to 350 meters is very fast compared to distances above 350 meters. Similar results were obtained for PM2.5 as well. It indicates that the PM monitoring devices shall be deployed at most 350 meters apart to capture the spatial variability of PM accurately.



%
\section{Conclusion}
In this paper, an end-to-end low-cost IoT system is developed and densely deployed in Indian urban settings for monitoring PM with fine spatial and temporal resolution. For evaluating the dense deployment, 49 calibrated devices were deployed covering a $4\, \text{km}^2$ area in Hyderabad, the capital city of Telangana state and the fourth most populated city in India. For data visualization, a web-based dashboard was developed for the real-time interface of PM data. The measurements over the year clearly show a significant difference between the mean and variance of PM values across different locations and seasons. The mean values and the variance were significantly higher in winter than in the summer and the monsoon. The IDW-based spatial interpolation results in monsoon, winter and summer at three different times show significant spatial variations in PM10 values. Furthermore, variation in PM values before and after the bursting of firecrackers on the day of Diwali is clearly visible in the results. The results also show noticeable temporal variations, with PM10 values rising by 4-5 (AV64) times at the same spot in a few hours, coinciding with Diwali celebrations and identifying the hotspots in dense deployment, which is not noticeable in sparse deployment. It has been shown that the correlation coefficient among a set of devices in the area has low values demonstrating that the PM values across a small region may be significantly different. A 350 m distance has been estimated for optimal device deployment for this data set based on insights deduced from the correlation versus distance plot. Thus, there is a need for dense deployment to understand the effect of local pollutants in the air and for improved spatial and temporal resolution of the pollutant data.
\bibliographystyle{IEEEtran.bst}
\bibliography{ref}
\end{document}